\documentclass[structabstract]{aa}
\usepackage{graphicx}
\usepackage{amsmath}
\usepackage{txfonts}
\usepackage{natbib}
\begin{document}
   \title{The gravitational wave signal from diverse populations of double white dwarf binaries in the Galaxy}

   \subtitle{}

\titlerunning{Gravitational Waves from Double White Dwarfs}

    \author{Shenghua Yu
          \and
          C.~Simon Jeffery
          }

   \authorrunning{S.~Yu \& C.~S.~Jeffery}

   \institute{Armagh Observatory, College Hill, Armagh, BT61 9DG,
   Northern Ireland\\
              \email{syu@arm.ac.uk, csj@arm.ac.uk}
               \\
             }

   \date{Received ; accepted }


  \abstract
   {The gravitational wave (GW) background in the range 0.01 $-$ 30 mHz has been assumed to be dominated
     by unresolved radiation from double white dwarf binaries (DWDs). Recent
     investigations indicate that, at short periods, a number of DWDs
     should be resolvable sources of GW.}
   {To characterize the GW signal which would be detected by LISA from DWDs in the
Galaxy.}
   {We have constructed a Galactic model in which we consider distinct
     contributions from the bulge, thin disc, thick disc, and halo,
     and subsequently executed a population synthesis approach to
     determine the birth rates, numbers, and period distributions of
     DWDs within each component. }
   {In the Galaxy as a whole, our model predicts the current birth rate of DWDs to be $3.21\times10^{-2}$ yr$^{-1}$, 
    the local density to be 2.2$\times10^{-4}\rm pc^{-3}$ 
     and the total
     number to be $2.76\times10^{8}$.
     Assuming SNIa are formed from the merger of two CO white dwarfs,
     the SNIa rate should be 0.0013 yr$^{-1}$.
    The frequency spectra of DWD strain amplitude and number
     distribution are presented as a function of galactic
     component, DWD type, formation channel, and metallicity. }
   {
     We confirm that CO+He and He+He white dwarf
     (WD) pairs should dominate the GW signal at very high frequencies
     ($\log f\,{\rm Hz^{-1}} > -2.3$), while CO+CO and ONeMg WD pairs have a dominant
     contribution at $\log f\,{\rm Hz^{-1}} \leq -2.3$. Formation
     channels involving two common-envelope (CE) phases or a stable
     Roche lobe overflow phase followed by a CE phase dominate the
     production of DWDs detectable by LISA at $\log f\,{\rm Hz^{-1}} >
     -4.5$.
     DWDs with the shortest orbital periods will
     come from the CE+CE channel. The Exposed Core plus CE channel is a minor
     channel. A number of resolved DWDs would be detected, making up
     0.012\% of the total number of DWDs in the Galaxy. The majority
     of these would be CO+He and He+He pairs formed through the CE+CE
     channel.
    }

   \keywords{Gravitational waves --
                Stars: binaries: close -- Stars: white dwarfs -- Galaxy: structure
 -- Galaxy: stellar content
               }

   \maketitle
%

\section{Introduction}

Substantial efforts have been made to measure gravitational waves
that, according to Einstein, should permeate the whole Universe
\citep{Press72,Tyson78}. Today, four major ground-based detectors
are in operation (LIGO, VIRGO, GEO600, and TAMA300), but no direct
evidence for the existence of gravitational waves (GW) has been
established.

Following the proposed launch of LISA (Laser Interferometer Space
Antennae), it is anticipated that the most easily detected waves
will originate from close compact binary stars. In addition to
providing a direct test of general relativity, the observations of
GW from such systems will contribute to understanding close binary
star evolution, the distribution of X-ray sources, $\gamma$-ray
bursts and supernovae explosions, and may also improve knowledge of
galactic structure.

As a result of recent and current surveys, increasing numbers of
candidate compact binaries are being found, including double neutron
stars (NS) and double white-dwarf (DWD) systems. AM\,CVn variables
are thought to be semi-detached ultra-compact binary stars systems in which a
white dwarf accretes from a degenerate (or semi-degenerate)
companion with an orbital period of less than 80 minutes. Currently
the number of known AM\,CVn systems is approximately 20
\citep{Ramsay07,Roelofs07}, while those of detached DWD and NS
systems with known orbital periods are 24 and 8 respectively.

Theoretical studies of the dominant GW signal in the Galaxy have
been pursued for several decades. \citet{Mironovskii66} calculated
the spectral density and total flux due to GW from W\,UMa binaries
and predicted a very low flux with a spectrum peaked at a period of
about 0.17\,d (0.07\,mHz) (A- and W-type W\,UMa binaries have
orbital periods in the range 0.4 -- 0.8\,d and 0.22 -- 0.4\,d,
respectively). Subsequently \cite{Evans87} demonstrated that
degenerate dwarf binaries should be detectable sources of
gravitational waves. More recent work indicates that
DWD binaries are crucial GW sources in the
0.1--100\,mHz range
\citep{Hils90,Hils00,Nelemans01b,Nelemans04,Farmer03,Willems07}. The
results of \citet{Ruiter09} show the halo signal would have a
negligible effect on the detection of LISA sources in comparison
with that of the disc and bulge.

Such population synthesis models are sensitive to the
adopted inputs, including the ¡°fast¡± stellar evolution models, the
Galactic model, including mass and metallicity distribution, initial
mass function, and star formation rate, and various parameters
governing binary star evolution, such as mass loss and exchange
in binary stars. It is important to ensure that previous
results are robust by reproducing them in independent calculations,
and to identify the major areas of uncertainty in these
models. It is also important to validate a new code by testing it
against established benchmarks, before using it to explore new
physics. This paper addresses both of these functions.

To discuss the GW background due to a population of galactic DWDs,
we have executed a simulation based on a population synthesis
approach for DD binaries which uses a multi-component model of the
Galaxy. We do not include the contribution of DD's containing an NS 
or black hole (BH) since the formation and evolution of these await
further theoretical and observational constraint. Other parameters
of our stellar evolution model are taken from the best model of
\citet{Han98}.

In addition to the GW signal, our model provides an estimate of the
current birth rate, total numbers and properties ($e.g.$ the primary
mass and orbital period) of DWDs which may be compared with
observation.

In \S2 we describe the model for generating a GW signal from a
binary, the stellar evolution model for generating a population of
compact binaries, and the model for generating their galactic
distribution. In \S3, we show the main results from these models in 
terms of the overall Galactic GW signal, and in terms of the contribution 
from various components of the DWD population. Section 4
compares these results with previous work, and conclusions are drawn in \S5.


\section{Simulations}
\label{sec_model}

\subsection{The GW signal from a binary star}

The power of the GW radiated from two point masses in the $n^{\rm th}$ harmonic 
is given by
\begin{equation}
L_{\rm GW}(n,e)=
\frac{32}{5}\left(\frac{G^{4}}{c^{5}}\frac{\mu^{2}M^{3}}{a^{5}}\right)g(n,e)
\label{eq_lgw}
\end{equation}
where $G$ is Newton's gravitational constant, $c$ is the speed of
light, $a$ is their orbital separation, $e$ is the eccentricity of 
the orbit, $M$ and $\mu$ are the total and reduced masses of the
system, and $g(n,e)$ is a coefficient expressing the relative power in
each harmonic and is given by \citet{Peters63}. 
If $m_{1}$ and $m_{2}$ are the masses of the components, then
\begin{equation}
M=m_{1}+m_{2},~~\mu=m_{1}m_{2}/M.
\end{equation}

The GW energy flux density incident on the Earth is
\begin{equation}
\frac{L_{\rm GW}}{4\pi d^{2}}=\frac{c^{3}}{16\pi
G}<\dot{h}_{+}^{2}+\dot{h}_{\times}^{2}>,
\label{eq_fgw}
\end{equation}
where $d$ is the distance, $h_{+}$ and $h_{\times}$ are
dimensionless functions for  the amplitude of the wave in two
orthogonal polarizations, and dots denote time derivatives
\citep{Isaacson68,Press72,Evans87}. Assuming a binary system in a
circular orbit of period $P_{\rm orb}$ at a distance $d$, $h_{+}$
and $h_{\times}$ are given by
\begin{equation}
h_{+}=\frac{G^{5/3}}{c^{4}}\frac{1}{d}2(1+\textrm{cos}^{2}i)(\pi
f_{\rm GW}M)^{2/3}\mu \textrm{cos}(2\pi f_{\rm GW}t),
\label{eq_hplus}
\end{equation}
\begin{equation}
h_{\times}=\pm\frac{G^{5/3}}{c^{4}}\frac{1}{d}4\textrm{cos}i(\pi
f_{\rm GW}M)^{2/3}\mu \textrm{sin}(2\pi f_{\rm GW}t),
\label{eq_htimes}
\end{equation}
where $f_{\rm GW}=2/P_{\rm orb}$ is the frequency of the emitted GW,
and $i$ is the inclination angle ($i=90^{\circ}$ indicates 
a binary which is viewed edge-on) \citep{Landau75}. 

The so-called {\it strain amplitude} $h$, defined as
$h^{2}=\frac{1}{2}[h_{+,max}^{2}+h_{\times,max}^{2}]$, describes the
strength of the signal to be measured by GW detectors. From
Eqs.~\ref{eq_fgw}, \ref{eq_hplus} and \ref{eq_htimes}, we obtain
\begin{equation}
h=\left(\frac{4GL_{\rm GW}}{\pi c^{3}f_{\rm GW}^{2}4\pi d^{2}}\right)^{1/2}
\label{eq_strain}
\end{equation}
\citep{Douglass79,Nelemans01b}. 
So from Eqs.\ref{eq_lgw}, \ref{eq_strain},  and 
Kepler's law ${a^{3}}/{P_{\rm orb}^{2}}=G(m_{1}+m_{2})/4\pi^{2}$, we
can write $h(n,e)$ in the $n^{\rm th}$ harmonic as
\begin{equation}
\begin{split}
h(n,e)& =
1.0\times10^{-21} \\
& \times \left(\frac{g(n,e)}{n^{2}}\right)^{1/2}
\left(\frac{\mathcal{M}}{M_{\odot}}\right)^{5/3} \left(\frac{P_{\rm
orb}}{\rm h}\right)^{-2/3} \left(\frac{d}{\rm kpc}\right)^{-1}.
\end{split}
\label{eq_hne}
\end{equation}
Equation~\ref{eq_hne} gives the strain amplitude in the $n^{\rm th}$
harmonic with a frequency $f_{\rm GW}^n = n / P_{\rm orb}$, and
where $\mathcal{M}\equiv\mu^{3/5}M^{2/5}$ is the so-called {\it
chirp mass}. Especially for a circular orbit, {\it i.e.} 
$e=0,~ n=2$, we have 

\begin{equation}
\begin{split}
h(n,e) & = h(2,0) = 5.0\times10^{-22} \\
& \times \left(\frac{\mathcal{M}}{M_{\odot}}\right)^{5/3}
\left(\frac{P_{\rm orb}}{\rm h}\right)^{-2/3} \left(\frac{d}{\rm
kpc}\right)^{-1}.
\end{split}
\label{eq_hneco}
\end{equation}

\subsection{Stellar evolution models}

A binary star population synthesis calculation requires the
predicted evolution of many thousands of binary stars, only a
fraction of which yield systems of interest (\S \ref{sec_formation}). 
We have obtained this predicted evolution using the fast binary star 
evolution (BSE) code  described by \citet{Hurley00,Hurley02} and previously
referred to by \citet{Han98}. However, we have modified the code
in a number of ways, including several of the approximation formulae 
and the treatment of the common envelope ejection mechanism.
This section describes these changes, including the approximations 
used to simulate changes in mass and angular momentum due to the 
interaction of binary star components. 

\subsubsection{Common-envelope (CE) evolution}
\label{sec_ce}

RLOF occurs in binary systems when one star fills its Roche lobe
either by evolutionary expansion or by orbital shrinkage. The Roche
radius of the primary is given by
\begin{equation}
\frac{R_{\rm L_{1,2}}}{a}=\frac{0.49q_{1,2}^{2/3}}{0.6q_{1,2}^{2/3}+\rm
ln(1+\it q_{\rm 1,2}^{\rm 1/3} )} \label{eq_rocherad}
\end{equation}
where $a$ is more generally the semi-major axis of the orbit and
$q_{1,2}=m_{1,2}/m_{2,1}$ \citep{Eggleton83}.
Studies of the critical value for stable RLOF have shown that
$q_{\rm c}$ is a function of primary mass $m_{1}$, its core mass
$m_{1\rm c}$, and mass-transfer efficiency of the donor. We adopt
\begin{equation}
q_{\rm c}=\left(1.67-x+2\left(\frac{m_{1\rm
c}}{m_{1}}\right)^{5}\right)/2.13
\end{equation}
where $x$ = 0.3 is the exponent of the mass-radius relation at
constant luminosity for giant stars \citep{Hurley00,Hurley02}.

The calculation of the properties (e.g. orbital separation) of
a binary after CE ejection in our model is based on the variation of angular
momentum ($\gamma$-mechanism). In the case of non-conservative mass
transfer, the angular momentum loss of a binary system would be
described by the decrease of primary mass times a $\gamma$ factor
\citep{Paczynski67a,Nelemans00}:
\begin{equation}
\frac{J_{\rm i}-J_{\rm f}}{J_{\rm i}}=\gamma \frac{m_{1}-m_{1\rm
c}}{M},
\label{eq_am1}
\end{equation}
where $J_{\rm i}$ is the orbital angular momentum of
the pre-mass transfer binary, and $J_{\rm f}$
is the final orbital angular momentum after CE ejection.
If initial and final orbits are assumed circular, the fraction of angular 
momentum lost during the mass transfer, $J_{\rm i} - J_{\rm f}$, becomes

\begin{equation}
J_{\rm i} - J_{\rm f} \approx \frac{m_{1}m_{2}}{M}a_{\rm i}^{2}\omega_{i} 
   - \frac{m_{\rm 1c}m_{2}}{(m_{\rm 1c}+m_{2})}a_{\rm f}^{2}\omega_{f}
\label{eq_am2}
\end{equation}
where $\omega_{i}$ and $\omega_{f}$ are the circular frequency of the binary 
before and after CE ejection. Combining \ref{eq_am1} and \ref{eq_am2} with 
Kepler's law, we have the ratio of final to initial orbital separation

\begin{equation}
\frac{a_{\rm f}}{a_{\rm i}}=\left(\frac{m_{1}}{m_{1\rm
c}}\right)^{2}\left(\frac{m_{1\rm c}+m_{2}}{M}\right)\left(1-\gamma \frac{m_{1}-m_{1\rm
c}}{M}\right)^{2}.
\end{equation}
\citet{Nelemans05} investigated the mass-transfer phase of the
progenitors of white dwarfs in binaries employing the
$\gamma$-mechanism based on 10 observed systems and suggested the
value of $\gamma$ in the range of 1.4 $\sim$ 1.7. We here adopt
$\gamma=1.5$, following the recommendation by \citet{Nelemans00}.

In order to eject a binary star common envelope, 
conservation of energy requires that the envelope binding energy, 
including gravitational binding and recombination energies,  
must be equal to the orbital energy \citep{Webbink84,Webbink08}. The ratio 
of final to initial orbital separation can be expressed as 
\begin{equation}
\frac{a_{\rm f}}{a_{\rm i}}=\frac{m_{1\rm c}}{m_{1}}\left[1+2\left(\frac{1}{\beta_{\rm CE}\lambda_{\Omega}R_{\rm L1}}-
\frac{1}{\lambda_{\rm P}R_{L1}}-\frac{\chi_{\rm eff}a_{\rm i}}{Gm_{1}}\right)\left(\frac{m_{1}-m_{1\rm c}}{m_{2}}\right)\right]^{-1},
\end{equation}
where the coefficients are given by \citet{Webbink08}. 

Both CE ejection mechanisms can reproduce observations \citep{Nelemans00,Nelemans05,
Webbink08}. We here adopt the $\gamma$-algorithm in order to reduce the number of 
free parameters in the model. 

We note that there is a major difference between the $\alpha-$ and
  $\gamma-$ algorithms. In both, the first CE phase can lead to the
  production of a low-mass binary with $q \approx 1$. However, the
  $\alpha-$ algorithm requires a significant spiral-in stage in order
  to eject the envelope whilst the $\gamma-$ mechanism does not,  
  implying that the orbital separation can be larger after CE ejection 
  in the latter case \citep{Nelemans00,Webbink08}. 

\subsubsection{Stellar wind and accretion}
\label{windandaccretion}

We assume that the existence of a close companion will increase the
rate of mass-loss ($\dot{m}$) due to a stellar wind from a star in a
binary system. The tidal enhancement of $\dot{m}_{1}$ is modelled by
\citet{Reimers75} formula with an extra tidal term by
\citet{Tout88}:
\begin{equation}
\dot{m}_{1}=-4\times10^{-13} \frac{r_{1}l_{1}}{m_{1}}
 \left( 1 + B\times
\left( {\rm min}(\frac{\it r_{\rm 1}}{\it r_{\rm
L_{1}}},\frac{1}{2}) \right)^6  \right) {\rm yr^{-1}},
\label{eq_mdot}
\end{equation}
where $r_{1}$, $l_{1}$ and $m_{1}$ are stellar radius, luminosity
and mass in solar units, respectively. We take $B=1000$
\citep{Han98}.

Some of the mass lost in the wind may be accreted by the companion.
The classical accretion rate formula is given by \citet{Bondi44},
and the mean accretion rate \citep{Boffin88} is
\begin{equation}
\dot{m}_{2}=\frac{-1}{\sqrt{1-e^{2}}} \left(
\frac{Gm_{2}}{v^{2}_{\rm w}} \right)^{2}
       \frac{\beta_{\rm w}}{2a^{2}}
       \frac{1}{(1+v^{2})^{3/2}}\dot{m}_{1},
\end{equation}
where $1 \leqslant \beta_{\rm w} \leqslant 2$ is an arbitrary
parameter ($\beta_{\rm w}= 3/2$, \citep{Boffin93}), $v_{\rm w}$ is
the wind velocity, $ v^{2}= v^{2}_{\rm orb} / v^{2}_{\rm w}$, and
$v_{\rm orb} = (GM/a)^{1/2}$ is the orbital velocity. Here, we adopt
$v_{\rm w} = 20 {\rm km s^{-1}}$. The angular momentum of a star 
will change if the star loses or gains mass. We assume that the specific 
angular momentum in the wind is proportional to the angular velocity
at the surface of the mass-losing star, and that the wind 
angular momentum is transferred with 100\% efficiency 
\citep{Hurley02}.

\subsubsection{Gravitational radiation and magnetic braking}

A close compact binary system driven by gravitational radiation would
eventually undergo a mass transfer phase, ultimately leading to
coalescence. The orbital changes due to gravitational radiation from
two point masses are predicted to be \citep{Hurley02}:
\begin{equation}
\begin{split}
\frac{\dot{a}}{2a}=\frac{\dot{J}_{\rm gr}}{J_{\rm orb}} =&
 -8.315\times10^{-10}
 \left(\frac{m_{1}}{M_{\odot}}\right)
 \left(\frac{m_{2}}{M_{\odot}}\right)
 \left(\frac{M}{M_{\odot}}\right)
 \left(\frac{a}{R_{\odot}}\right)^{-4}\\
& \times \left(\frac{1+\frac{7}{8}e^{2}}{(1-e^{2})^{5/2}}\right)
{\rm yr^{-1}} ,
\end{split}
\end{equation}

\begin{equation}
\begin{split}
\frac{\dot{e}}{e} = &
 -8.315\times10^{-10}
 \left(\frac{m_{1}}{M_{\odot}}\right)
 \left(\frac{m_{2}}{M_{\odot}}\right)
 \left(\frac{M}{M_{\odot}}\right)
 \left(\frac{a}{R_{\odot}}\right)^{-4}\\
 & \times \left(\frac{\frac{19}{6}+\frac{121}{96}e^{2}}{(1-e^{2})^{5/2}}\right)
 {\rm yr^{-1}}.
\end{split}
\end{equation}

Gravitational radiation could explain the formation of cataclysmic
variables (CVs) with orbital periods less than 3h, while magnetic
braking of the tidally coupled primary by its own magnetic wind
would account for orbital angular-momentum loss from CVs with
periods up to 10\,h \citep{Faulkner71,Zangrilli97}. We use the
formula for the rate of angular-momentum loss due to magnetic braking
derived by \citet{Rappaport83} and \citet{Skumanich72}:
\begin{equation}
\dot{J}_{\rm mb}=-5.83\times10^{-16} 
\frac{m_{\rm env}}{m}
\left( \frac{r\omega_{\rm spin}}{\rm R_{\odot} yr^{-1}} \right)^{3}
~{\rm M_{\odot}R_{\odot}^{2}yr^{-2}},
\end{equation}
where $r$, $m_{\rm env}$ and $m$ are the radius, envelope mass and mass of 
a star with a convective envelope, and $\omega_{\rm spin}$ is 
the spin angular velocity of the star. 

\subsubsection{Tidal interaction}

Observations indicate that stellar rotation in a close binary system 
tends to synchronize with the orbital motion \citep{Levato74,Strassmeier96}. 
The two tidal dissipation processes, {\it i.e.} turbulent dissipation of 
the equilibrium tide and radiative damping of the dynamical tide, are able 
to account for the observed orbital circularization of close binary and 
components rotating in synchronism with orbital motion; see \citet{Zahn05} 
for a review. The calculation of tidal interaction in our model follows
\citet{Hurley02} and adopts an equilibrium tide for the stars with a 
convective envelope \citep{Rasio96} and a dynamical tide for the stars
with a radiative envelope \citep{Zahn75,Zahn77}. 

In addition, we note that a hydrodynamical mechanism may account for 
the orbital circularization as it is concomitant with the hydrodynamical spin-down 
of the components both in early-type and late-type detached close binaries 
\citep{Tassoul88}.
Since this mechanism would be more effective than the tidal
interaction at 
circularizing the orbit of a binary, we adopt the tidal dissipation as
an upper limit to the circularization 
timescale.

The tidal synchronisation
time is then
\begin{equation}
t_{\rm synch}=1.3\times10^{7}q_{2}^{2}
\left(\frac{m_{2}/M_{\odot}}{l_{2}/L_{\odot}}\right)^{5/7}
\left(\frac{r_{2}}{a}\right)^{-6} {\rm yr}
\end{equation}
for a tide raised on a white dwarf secondary of mass $m_{2}$
\citep{Campbell84}, where $r_{2}$, $l_{2}$ and $q_{2}$ represent its radius, 
luminosity and mass ratio respectively. $t_{\rm synch}$ would be a lower limit if
tidally-induced non-radial oscillations were included. The orbit of DWDs 
should be circularized already, and a circularization time-scale is not 
relevant. However, the synchronization time-scale is essential as the 
companion may be spun up by mass transfer \citep{Hurley02}. 

\subsubsection{Mass transfer in compact binaries}

As a result of gravitational radiation, mass transfer or direct
merger between two compact binaries is inevitable if the compact
object with lower mass fills its Roche lobe. The analytic
mass-radius relation for zero-temperature white dwarfs, is
\citep{Nauenberg72}
\begin{equation}
r=0.01125R_{\odot}(m'^{-2/3}-m'^{2/3})^{1/2},
\end{equation}
where $m'=m/1.433M_{\odot}$ is the mass in terms of the
Chandrasekhar mass. Combining with Eq.
~\ref{eq_rocherad}, we deduce an equilibrium mass transfer rate
\begin{equation}
\dot{m}_{2}=\frac{C_{2}L_{\rm Edd}/\phi_{\rm r1}-2m_{2}/\tau_{\rm
GR}}{\xi_{\rm ad}-C_{1}-C_{2}\phi_{\rm L2}/\phi_{\rm r1}},
\label{eq_masstransfer}
\end{equation}
where $L_{\rm Edd}=4\pi r_{2}^{2}cg / \kappa $ is the Eddington
luminosity, $\kappa=0.2(1+X)\,{\rm cm^{2}g^{-1}}$ is the opacity of 
the accreted gas, assumed here to be due to electron scattering, $X$ is the hydrogen mass fraction, $g$
is the gravitational acceleration, and other coefficients are as given
by \citet{Han99}. $\tau_{\rm GR}$ is the time scale
for orbital angular-momentum loss due to gravitational radiation
\citep{Landau58} with
\begin{equation}
\tau_{\rm
GR}=\frac{5}{32}\frac{c^{5}}{G^{3}}\frac{a^{4}}{m_{1}m_{2}M}.
\label{eq_grtime}
\end{equation}
and $C_1$, $C_2$ $\xi_{\rm ad}$ and $\phi_{\rm L1}$ as given by
\citet{Han99}.

For stable mass transfer, \citet{Han99} give the critical mass ratio
\begin{equation}
q = \frac{m_{2}}{m_{1}} < q_{\rm c}\approx 0.7-0.1(m_{1}/M_{\odot}). \label{eq_qc}
\end{equation}
Here, $m_{2}$ is donor.
If $q \gtrsim q_{\rm c}$, the mass transfer becomes dynamically unstable,
eventually causing a runaway merger.

\citet{Nelemans01a} give an alternative value $q_{\rm c} \approx
5/6+\zeta(m_{2})/2$, $\zeta\equiv {\rm d} \ln r / {\rm d} \ln m$,
using the size of the Roche lobe and the mass transfer rate derived
by \citet{Paczynski67b}. We adopt the \citet{Han99} approximation.

Finally, we give a formula for the Eddington accretion rate
\citep{Cameron67,Hurley02}
\begin{equation}
\dot{M}_{\rm Edd} = 2.08\times10^{-3}(1+X)^{-1}r_{1} {\rm
M_{\odot}yr^{-1}},
\label{eddington}
\end{equation}
which places a significant limit on the amount of mass accreted by
compact objects. 

We set this limit only for the accretion of 
compact objects. However, whether the Eddington limit should be applied 
or not needs to be further investigated, since the luminosity generated 
by accretion in excess of this limit might be carried away from the system via a 
strong wind, or transfered to a disc. Super-Eddington accretion rates may 
be important for the formation of low-mass X-ray binaries and millisecond 
pulsars \citep{Webbink97} and also for modeling X-ray emission from quasars 
due to accretion from a disc onto a fast rotating black hole 
\citep{Beloborodov98}. Imposing the limit diminishes the birth rate of 
type Ia supernovae \citep{Livio00}. 

\subsubsection{The formation channels for compact binaries}
\label{sec_formation}

One can find a detailed discussion of the formation channels 
of compact binaries in \citet{Han98}. 
We here only summarize the three main evolution channels which produce 
the majority of compact binaries : \\
I:~ Stable RLOF $+$ CE ejection; \\
II:\, CE ejection $+$ CE ejection; \\
III: Exposed core $+$ CE ejection.

In channel I, a binary with mass ratio $q=m_1/m_2$ less than some
critical value $q_{\rm c}$ will experience dynamically stable mass
transfer if the primary fills its Roche lobe while the star is in
the Hertzsprung gap or on the red giant branch. The
primary will become a compact object and the orbital separation will
change as
\begin{equation}
-{\rm d} \ln a =2 {\rm d} \ln m_{\rm 2}+ 2\alpha_{\rm RLOF} {\rm d}
\ln
 m_{\rm 1} + {\rm d} \ln (m_{\rm 1}+m_{\rm 2})
\end{equation}
where $\alpha_{\rm RLOF}$ is the mass-transfer efficiency for stable
Roche-lobe overflow (RLOF) \citep{Han95}. Here, we take $\alpha_{\rm
RLOF}=0.5$ \citep{Paczynski67a,Refsdal74}. Subsequently, if the {\it
secondary} fills its Roche lobe while it is in the Hertzsprung gap or 
on the red giant branch, then RLOF will occur. 

If the adiabatic response of the radius 
of the mass donor is less than the change of its Roche lobe radius 
with respect to a change of mass, {\it i.e.} 
$\left(\frac{\partial \ln R_{\rm donor}}{\partial \ln M_{\rm donor}}\right)_{\rm ad}<
\left(\frac{\partial \ln R_{\rm RLOF}}{\partial \ln M_{\rm donor}}\right)_{\rm RLOF}$, 
mass transfer will be unstable and a common envelope (CE) will form.
Interaction (friction) between the compact cores and the CE will
convert orbital energy into kinetic energy, heating and expanding
the CE. If the energy conversion mechanism is sufficiently
efficient, the CE will be expelled and a compact binary with a short
orbital period will result.

Channel II demands that the binary has $ q > q_{\rm c}$ so that, if
the primary fills its Roche lobe while in the Hertzsprung gap or on 
the red giant branch,
unstable mass transfer  will lead to the formation of a CE.
Similarly, if sufficient orbital energy can be extracted, the CE
will be ejected to leave a binary containing a compact object.
Following evolution of the secondary away from the main sequence, 
a second CE could form and be ejected to leave a compact
binary.

Channel III represents a variation of channels I and II in which the
envelope of a massive primary is removed by a stellar wind rather
than a first CE ejection. CE ejection following evolution of the
secondary may also give rise to a compact binary.

In summary, CE plus CE represents the major channel. Up to 
74\% \citep{Han98}, or 67\% in our model (see Table \ref{tab_type}), 
of compact binaries may be generated through this channel depending on the choice of 
parameters. This may also be the most efficient channel to 
produce compact binaries with very short orbital periods. The stable 
RLOF plus CE channel can account for up to 50\%  of compact binaries 
\citep{Han98}, or 30\% in our model. Exposed core plus CE is a minor
channel which could make a significant contribution when a binary evolves with a 
strong tidally-enhanced stellar wind. Both the results of \citet{Han98} 
and our own show this channel to be negligible for compact binaries with 
orbital periods shorter than 70 hr.

\begin{table}
\begin{center}
\caption{Density laws and associated parameters. $r$ is the
spherical radius from the center of the Galaxy and $r_{0}$ is bulge
scale length; $R$ and $z$ are the natural cylindrical coordinates of
the axisymmetric disc, $h_{R}$ is the scale length of the disc,
$h_{z}$ is the scale height of the thin disc, $h'_{z}$ is the scale
height of the thick disc; $a$ is the radius of the halo and $a_{0}$
is a constant; $\rho_{c}$ is the central mass density. \label{tab_densitylaws}}
\begin{tabular}{lccc}
\hline
&density law & constants   & $\rho_{c}$  \\
&            &  (kpc) & (M$_{\odot}\rm pc^{-3}$) \\
\hline
Bulge &$e^{-(r/r_{0})^{2}}$ &$r_{0}=0.5$ & $\frac{M_{\rm b}}{4\pi r_{0}^{3}}=12.73$ \\
\\
Thin disc &$e^{-R/h_{R}}\textrm{sech}^{2}(-z/h_{z})$ &$h_{R}=2.5$   & $\frac{M_{\rm tn}}{4\pi h_{R}^{2}h_{z}}=1.881$ \\
          &                                          &$h_{z}=0.352$ & \\
\\
Thick disc &$e^{-R/h_{R}}e^{-z/h'_{z}}$ & $h_{R} = 2.5$  & $\frac{M_{\rm tk}}{4\pi h_{R}^{2}h'_{z}}=0.0286$\\
           &                            & $h'_{z}=1.158$ & \\
\\
Halo &$ [(1+(\frac{a}{a_{0}})^{2})]^{-1}$ &$a_{0} = 2.7$ & 0.108\\
\hline
\end{tabular}
\end{center}
\end{table}

\begin{figure}
\centering
\includegraphics[width=9.6cm,clip,angle=0]{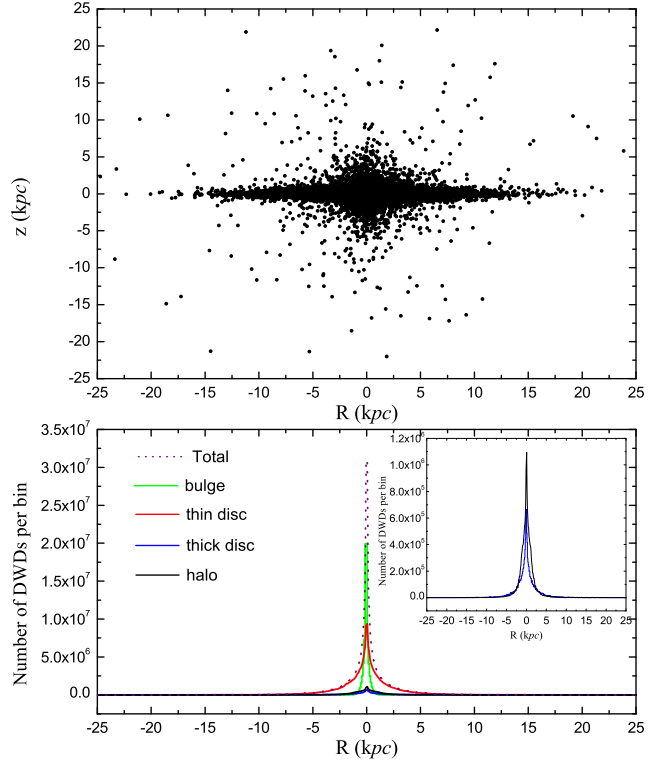}
\caption{The top panel shows the distribution of stars in the 
R(Galactic plane)-z(height) diagram for our Galactic model at age 10 Gyr, 
using a random sample of $\sim70,000$ stars. The lower panel gives 
the number of DWDs in the Galaxy as a function of galactic radius, 
using a bin size of 0.1 k$pc$. The inset panel shows the thick disk and
halo distribution. We assume there is no immigration between each 
Galactic component, {\it i.e.} no mass transfer, no angular momentum transfer, 
no collision.}
\label{appfig}
\end{figure}

\subsection{Galactic structure}

A detailed model of the Galaxy is important in order to describe the
overall distribution of white dwarf binary systems, including
the distance of each from the Sun. It is believed that the current
Galaxy mainly comprises the bulge, the thin disc, thick disc and the
halo. We summarize our approximation for the Galactic density
distribution in Table~\ref{tab_densitylaws}.

Figure~\ref{appfig} shows the distributions of distances to the Earth
of compact binaries in the Galaxy and the bulge plus the thin disc.
We assume that the position of the sun is $R_{\rm sun} = 8.5$\,kpc
and $z_{\rm sun}$ = $16.5$\,pc \citep{Freudenreich98}.

(1) We adopt a normal density distribution for the spherical bulge
with a cut-off radius of 3.5\,kpc \citep{Nelemans04},
\begin{equation}
\rho_{\rm b}(r) = \frac{M_{\rm b}}{4\pi r_{0}^{3}} e^{-(r/r_{0})^{2}}
    \quad {\rm M_{\odot} pc^{-3}},
\end{equation}
where $r$ is the radius from the center of the Galaxy, 
$r_{0}= 0.5$\,kpc is the bulge scale length, and 
$M_{\rm b}$ is the mass of bulge (see \S \ref{sec_mcs} find the value). 
\citet{Robin03} suggest that the structure of the inner bulge
($<1^{\circ}$ from the Galactic center) is not yet well constrained
observationally. Consequently we here focus on the outer bulge and
make no allowance for any {\it additional} contribution to the
compact-binary population from the central region.

(2) A more complicated function is involved in the spatial density
distribution of stars in the disc. \citet{Sackett97} proposed a
double exponential distribution. \citet{Phleps00} derived three
functions for the star density distribution in their model of a thin
disc plus thick disc (exponential $+$ exponential, hyperbolic secant
$+$ exponential, and squared hyperbolic secant $+$ exponential,
respectively) from fits to deep star counts carried out in the Calar
Alto Deep Imaging Survey. \citet{Robin03} created a complicated 
function to construct the structure of the Galactic disc
which is in agreement with Hipparcos results and the observed
rotation curve. 

We here model the thin and thick disc components of the
Galaxy using a squared hyperbolic secant plus exponential distribution
expressed as:
\begin{equation}
\rho_{\rm d}(R,z) = \frac{M_{\rm d}}{4\pi h_{R}^{2}h} e^{-R/h_{R}}\rho(z)
 \quad {\rm M_{\odot} pc^{-3}},
\end{equation}
where $R$ and $z$ are the natural cylindrical coordinates of the
axisymmetric disc, and $h_{R}=2.5$\,kpc is the scale length of the
disc, $h=h_{z}$ for the thin disc, $h=h'_{z}$ for the thick disc, 
and $M_{\rm d}=M_{\rm tn}$ is the mass of the thin disc; 
$M_{\rm d}=M_{\rm tk}$ is the mass of the thick disc (see \S \ref{sec_mcs} 
find the values). 
$\rho(z)$ is the distribution in $z$, with:
\begin{equation}
\rho(z) = \textrm{sech}^{2}(-z/h_{z})~{\rm (thin~disc)}
\end{equation}
and
\begin{equation}
\rho(z) = e^{-z/h'_{z}}~{\rm (thick~disc)},
\end{equation}
where $h_{z}=0.352$\,kpc is the scale height of the thin disc and
$h'_{z}=1.158$\,kpc is the scale height of the thick disc. We neglect
the age and mass dependence of the scale height.

(3) For the halo, we employ a relatively simple density distribution
which is consistent with \citet{Caldwell81} and \citet{Robin03}: \footnote{Although a
distribution
  following $e^{-a^{1/4}}$ or $a^{1/4}$ might be a better choice
\citep{deVaucouleurs53, deVaucouleurs58}, we have chosen the form
given here for simplicity.}
\begin{equation}
\rho_{\rm h}(a) =  \rho_{\rm c_h} \times
\left(1+\left(\frac{a}{a_{0}}\right)^{2}\right)^{-1},
\end{equation}
where $a$ is the radius of the halo, $\rho_{\rm c_h}~=~0.108~M_{\odot}\rm pc^{-3} $
and $a_{0}~=~2.7~{\rm kpc}$.

\subsection{Population synthesis parameters in the Monte Carlo approach}
\label{sec_mcs}

In order to obtain a sample of compact binaries in the
Galaxy, we have performed a Monte Carlo simulation in which we need
five physical inputs:

(i) We assume that the star formation rate (SFR) in the bulge and
thin disc is the combination of a main star forming process (the first
item of the following function) and a minor star formation (the
second item of the function),
\begin{equation}
{\rm SFR}(t) = 11 e^{-(t-t_0)/\tau}+0.12(t-t_{0})~{\rm
M_{\odot}yr^{\rm -1}}, t>t_0
\end{equation}
where $t$ is time since the halo was formed. Assuming the current
age of the Galaxy is 14 Gyr, $t_0=4$ Gyr defines the age of the
bulge and thin disc to be 10 Gyr and $\tau=9$ Gyr yields a current
SFR $=4.82~{\rm M_{\odot} yr^{-1}}$, 1.45 ${\rm M_{\odot} yr^{-1}}$
in the bulge and 3.37 ${\rm M_{\odot} yr^{-1}}$ in the thin disc, on
average. These values are consistent with \citet{Smith78}, 
\citet{Timmes97} and \citet{Diehl06}. We assume that 
$\textrm{SFR}(t)=0$ in the bulge and thin disc when $0<t<t_{0}$.

In terms of these assumptions, we infer that the combined
mass of thin disc and bulge approaches $7.2\times 10^{10}~ {\rm
M_{\odot}}$, slightly higher than the mass of $7.0\times 10^{10}~
{\rm M_{\odot}}$ reported by \cite{Klypin02}. Within this mass, the
bulge contains $M_{\rm b}=2.0\times 10^{10} ~{\rm M_{\odot}}$ and the
remaining $M_{\rm tn}=5.2\times 10^{10}~ {\rm M_{\odot}}$ is in the thin disc.

We suppose a burst of star formation, effectively a $\delta$ function, 
happened at $t = 0$ Gyr for the halo, and at $t = 3$ Gyr for the thick disc 
\citep{Robin03} and no star formation thereafter. We assume that the
thick disc and the halo attain baryonic masses of $M_{\rm tk}=2.6\times
10^{9}~{\rm M_{\odot}}$ (5\% of thin disc) and $M_{\rm h}=1.0\times 10^{9}~{\rm
  M_{\odot}}$ respectively. These numbers are adopted for simplicity
solely to estimate their contribution to the GW signal.

(ii) The initial mass function (IMF) can be constrained by the local 
luminosity function, stellar density and potential. We here adopted the
IMF for the Galactic components based on the results of \citet{Robin03} and 
\citet{Kroupa93} constrained by the observations of \citet{Wielen83}, 
\citet{Popper80} and the Hipparcos mission \citep{Creze98,Jahreiss97}. 

For the bulge, we suppose an IMF following \citet{Robin03},
\begin{equation}
\xi(m)~\propto~m^{-2.35},~~m>0.7 {\rm M_{\odot}}
\end{equation}
where $m$ is the primary mass and $\xi(m)\textrm{d}m$ is
the number of stars in the mass interval $m$ to $m+\textrm{d}m$.

For the thin disc, we adopt the IMF of \citet{Kroupa93}\footnote{
\citet{Kroupa01} gives $\xi(m) \propto m^{-2.3}, 0.5 < m/{\rm M_{\odot}} $, 
see \S\ref{error} } which is similar to that of 
\citet{Miller79} and \citet{Zoccali00}. 
The primary mass is generated using the following formula
\begin{equation}
\xi(m)=\left\{
\begin{array}{c}
~0.035 m^{-1.3}~,~~0.08<m/{\rm M_{\odot}}<0.5,\\
0.019 m^{-2.2}~,~~0.5<m/{\rm M_{\odot}}<1.0,~\\
~~0.019 m^{-2.7}~,~~1.0<m/{\rm M_{\odot}}<100.0.
\end{array}
\right.
\label{eq_imftn}
\end{equation}

To the thick disc and the halo, we also apply a simple IMF of power
law,
\begin{equation}
\xi(m)~\propto~m^{-\alpha}. \label{eq_halo_imf}
\end{equation}
Here, we take $\alpha=$1.5 for the thick disc and the halo.

We have adopted a metallicity $Z=0.02$ (Population I) for the bulge,
thin disc and thick disc, and $Z=0.001$ for the halo. We have also
carried out calculations for the thin disc with metallicity $Z=0.001$
in order to see the effect of metallicity on the GW signal.

(iii)  We assume a constant mass-ratio distribution
\citep{Mazeh92,Goldberg94},
\begin{equation}
n(1/q) = 1,  0 \leqslant 1/q \leqslant 1,
\end{equation}

(iv) We employ the distribution of initial orbital separations used
by \citet{Han98} and \citet{Han03}, where they assume that all stars are
members of binary systems and that the distribution of separations
is constant in $\log a$ ($a$ is the separation) for wide binaries
and falls off smoothly at close separations:
\begin{equation}
a n(a)=\left\{
\begin{array}{c}
\alpha_{\rm sep}(\frac{a}{a_{0}})^{k},a\leqslant
a_{0},\\
\alpha_{\rm sep},a_{0}<a<a_{1}.
\end{array}
\right.
\label{eq_a}
\end{equation}
where $\alpha_{\rm sep}\approx0.070$, $a_{0}=10R_{\odot}$,
$a_{1}=5.75\times 10^{6}R_{\odot}=0.13$ pc, $k\approx1.2$. This
distribution implies that the number of binary systems per
logarithmic interval is constant. In addition, approximately 50 per
cent of all systems are binary stars with orbital periods of less
than 100 yr.

(v) The distribution of eccentricities of binaries follows
$P_{e}=2e$ \citep{Nelemans01b}.

\subsection{Rotation curve and local stellar density}

\begin{figure}
\centering
\includegraphics[width=10cm,clip,angle=0]{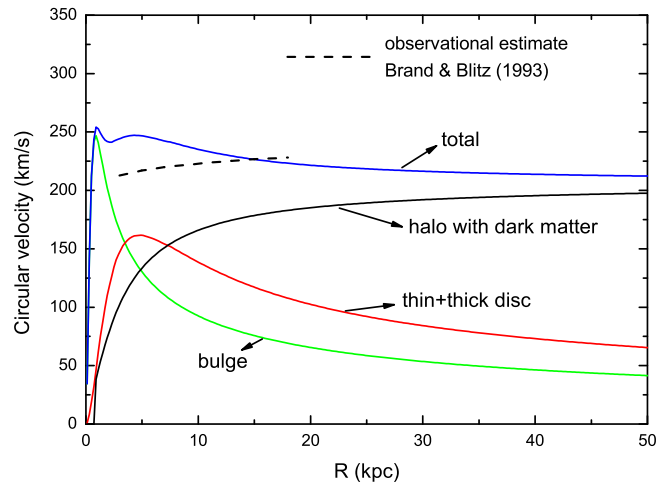}
\caption{ Circular velocity as a function of galactocentric distance $R$ 
from the Galatic model, showing the contribution due to different
components, 
{\it i.e.} the bulge, thin disc $+$ thick disc, and halo including
dark matter. 
The dashed line indicates the observational estimate by \citet{Brand93}. 
The spheroidal component due to the interstellar medium was not
considered separately.} 
 \label{fig_rotation}
\end{figure}

In order to see the influence of the Galactic model on the rotation curve 
of the Milky Way, we plotted Fig. \ref{fig_rotation}. We used the Miyamoto-Nagai potential 
\citep{Miyamoto75,Revaz09} in cylindrical coordinates for calculating the circular 
velocity of the bulge and disc components. For the dark matter halo, we adopted the 
potential of \citet{Caldwell81}. The observational estimate by
\citet{Brand93} is included for comparison. 

From the Galactic model, the total mass of the halo including 
dark matter is $4.5\times 10^{11}{\rm M_{\odot}}$ inside a sphere of 
radius 50\,kpc. In this paper, we only focus on the 
baryonic mass in the halo which is considered to be
$1\times10^{9}{\rm M_{\odot}}$. 
With the SFR adopted here, the baryonic mass in the bulge and disc
is at least $2\times10^{10}{\rm M_{\odot}}$ and $5.5\times10^{10} {\rm 
M_{\odot}}$ respectively, implying that our model requires no dark matter
component in the bulge or the thin disc.

Combining the Galactic model and the mass of the Galactic components, 
the stellar density in the solar neighbourhood is 
$0.064 {\rm M_{\odot}pc^{-3}}$, of which 
$6.27\times10^{-2}{\rm M_{\odot}pc^{-3}}$ is in the thin disc, 
$9.4\times10^{-4}{\rm M_{\odot}pc^{-3}}$ is in the thick disc, and 
$2.18\times10^{-5}{\rm M_{\odot}pc^{-3}}$ is in the halo. 
This is consistent with the Hipparcos result, $0.076\pm0.015 
{\rm M_{\odot}pc^{-3}}$ \citep{Creze98}. 
The local dark matter density in our model is about 
$0.01{\rm M_{\odot}pc^{-3}}$.

\subsection{Procedure}

In order to compute birth rates, number densities, space, mass and
orbital distribution of DWDs, we have adopted the following procedure.
For each Galactic component $g$ (bulge, thin disc, thick disc and
halo) having a total mass $M_g(t)=\int_{0}^{t} {\rm SFR}(t') {\rm d}t'$:
\begin{enumerate}
\item  Calculate a sample distribution of $k$ coeval binaries 
      having a total mass $m_{\rm p}$ and
      generated by the four Monte-Carlo simulation parameters $m$,
      $a$, $q$ and $e$.
\item Follow the evolution of each primordial binary to establish
      the properties of $n(t')$ DWDs formed from the original sample, and
      the number $x(t')$ of DWDs which merge, where $t'$ represents
      time since formation of each binary.
\item By combining $n(t')$ and $x(t')$ with the star formation rate, 
      compute the birth rate of DWDs $\nu(t) = \int_{0}^{t}
      (n(t')/m_{\rm p}) \times {\rm SFR}(t-t') {\rm d}t'$.
\item Compute the merger rate of DWDs $\xi(t) = \int_{0}^{t}
      (x(t')/m_{\rm p}) \times {\rm SFR}(t-t') {\rm d}t'$.
\item Evaluate the total number of DWDs $N(t) = \int_{0}^{t}
      \nu(t')-\xi(t') {\rm d}t'$.
\item For a given $t$ ({\it e.g.} 14 Gyr), use a second Monte Carlo
      procedure to generate the orbital properties and spatial
      distribution of $N(t)$   Galactic DWDs by interpolation and extrapolation on the
      $n$ DWDs from the simulation.
\item Sort the DWDs by orbital frequency.
\item Calculate the total strain amplitude $h^2$ from the number and
      distance of DWDs in each frequency bin.
\end{enumerate}

In our population synthesis simulation, we started with
$k = 1.20\times10^{7}$ primordial binaries ($4.0\times10^{6}$ in the bulge, 
$5.0\times10^{6}$ in the thin disc, $1.0\times10^{6}$ in the thick
disc and $2.0\times10^{6}$ in the halo) yielding a total of
$7.66\times10^{4}$ 
DWDs over all four components of the Galaxy at the present epoch, $t = 14\,$Gyr.

\begin{figure*}
\centering
\includegraphics[width=20cm,clip,bb=35 25 670 330,angle=0]{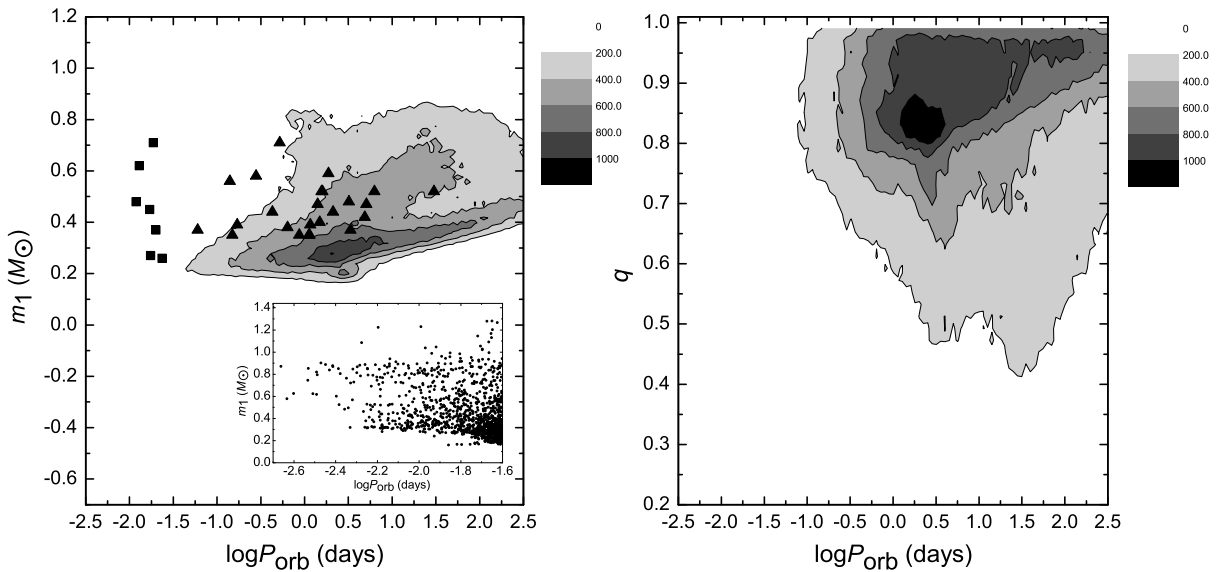}
\caption{Distribution of primary masses (left panel),
mass ratio $q=m_{2}/m_{1}$ (right panel) and orbital periods of DWDs in our model. 
The contour scale to the right of the figure represents the number of
systems in each bin. The present DWD population is plotted, but the total number of DWDs 
($2.76\times10^{8}$) is reduced by a factor of 100. 
Filled triangles and squares are for known WD+WD systems. 
The inset in the left panel shows the distribution
of primary mass and orbital periods at $\log P_{\rm orb} < -1.6$
days. Bin sizes are $\Delta \log (P_{\rm orb}/{\rm day})=0.05 $,
$\Delta m_1/{\rm M_{\odot}} = \Delta q = 0.02$.  }
\label{fig_massperiod}
\end{figure*}

\begin{table*}
\caption{Birth rates, local densities and numbers of DWDs. $\nu$ = current birth rate,
$\xi$ = current merger rate, $N$ = total number, $N_{\rm o}$ =
  the number for DWDs with a detectable component, 
  $X$ = merger number, $\rho_{\rm LD}$=local density of DWDs. The last  two rows represent the current birth rate of supernovae Ia 
  and the number of resolved DWDs detectable by LISA. Different thin disc
  models are denoted by $a$: $Z=0.02$; $b$: $Z=0.001$. Rates $\nu,\xi$
  are in yr$^{-1}$, $\rho_{\rm LD}$ in $pc^{-3}$. Note: a DWD is
    deemed detectable for $10^8$ yr after the second star became a
    WD. This definition follows \citet{Iben97} and the values are given
    for comparison with \citet{Han98}. }
\label{tab_birthrates}
\begin{center}
\begin{tabular}{lccccccc}
\hline
 &Bulge &Thin disc &Thick disc &Halo &Galaxy\\
 \hline
 $\nu-\xi$&$8.0\times10^{-3}$  &$^{a}2.1\times10^{-2}$& $1.4\times10^{-3}$ &$1.7\times10^{-3}$&$^{a}3.21\times10^{-2}$\\
      &                    &$^{b}2.7\times10^{-2}$&                   &                      &$^{b}3.81\times10^{-2}$\\
 $N  $&$7.6\times10^{7}$ &$^{a}1.7\times10^{8}$& $1.2\times10^{7}$ &$1.9\times10^{7}$&$^{a}2.76\times10^{8}$\\
            &                 &$^{b}2.3\times10^{8}$ &                 &                      &$^{b}3.37\times10^{8}$\\
 $N_{\rm o}$&$8.0\times10^{5}$ &$^{a}2.1\times10^{6}$& $1.4\times10^{5}$ &$1.7\times10^{5}$&$^{a}3.21\times10^{6}$\\
           &                  &$^{b}2.7\times10^{6}$&                  &                      &$^{b}3.81\times10^{6}$\\
 $\xi$&$4.0\times10^{-4}$ &$^{a}1.4\times10^{-3}$&$1.1\times10^{-4}$ &$8.3\times10^{-4}$& $^{a}2.74\times10^{-3}$\\
             &                  &$^{b}1.5\times10^{-3}$&                    &                     &$^{b}2.84\times10^{-3}$\\
 $X$  &$5.4\times10^{6}$ &$^{a}1.1\times10^{7}$&$1.2\times10^{6}$ &$1.2\times10^{6}$& $^{a}1.88\times10^{7}$\\
           &                  &$^{b}1.1\times10^{7}$&                  &                     &$^{b}1.88\times10^{7}$\\
 $\rho_{\rm LD}$  &$-$ &$^{a}2.1\times10^{-4}$&$4.4\times10^{-6}$ &$4.1\times10^{-7}$& $^{a}2.2\times10^{-4}$\\
           &                  &$^{b}2.8\times10^{-4}$&                  &                     &$^{b}2.9\times10^{-4}$\\
\hline
 \hline
 SNe Ia &                 &              &                    &                    &   $\nu=$ $^{a}1.3\times10^{-3}$     \\
 resolved &   14800              &    $^{a}18870  $        &                    &                    &  $^{a}N_{\rm r}=$ 33670     \\
          &                      &    $^{b}22670  $        &                    &                    &  $^{b}N_{\rm r}=$ 37470     \\
\hline
\end{tabular}
\end{center}
\end{table*}

\section{Results}

\subsection{Birth rates, local densities and numbers of DWDs}
\label{sec_birnum}

The results of our simulation are shown in
Table~\ref{tab_birthrates}, where we give the birth and merger rates
(number per year), local densities ($\rm pc^{-3}$) and total numbers of DWDs.
Figures are given for each component of the Galaxy and for
the Galaxy as a whole. Table~\ref{tab_birthrates} also shows the
numbers detectable if we assume that DWDs continue to be detectable
up to $10^{8}$ yr after their formation \citep{Iben97}. At the end
of Table~\ref{tab_birthrates}, we show the
birthrate for Type Ia supernovae $ = 0.0013$\,yr$^{-1}$,
assuming that all SNe\,Ia are formed from merging double
carbon-oxygen (CO) white dwarfs with total mass $>1.378 {\rm
  M_{\odot}}$ \citep{Martin06}.

The local densities of DWDs in our model are 
$2.1\times10^{-4}$, $4.4\times10^{-6}$, and   $4.1\times10^{-7} {\rm
    pc^{-3}}$ 
in the thin disc, thick disc, and halo, respectively. The predicted 
density of halo DWDs is significantly less than the observed
density of halo white dwarfs $=2\times10^{-4}{\rm pc^{-3}}$
\citep{Oppenheimer01}. 
This could be partially 
due to the star formation history, as we only have a single star 
burst of $1\times10^{9} {\rm M_{\odot}}$ at the formation 
of the halo. If we assume 10\% of halo WDs are in binary systems \citep{Holberg09}, 
the observation requires the local density of halo DWDs to be
$\approx2\times10^{-5}{\rm pc^{-3}}$. This would require 
the baryonic mass in the halo to be $5\times10^{10} {\rm M_{\odot}}$,
 11\% of the total mass of the halo, while the dark matter 
takes up the remaining 89\% of mass of halo. 

Table~\ref{tab_birthrates} also gives figures for the different thin
disc models discussed above. A low metallicity results in a
slight increase of the birth rate and number of WD binaries for a
given IMF, in line with the results of \citet{Han98}.

Figure~\ref{fig_massperiod} shows the distribution of orbital period $P_{\rm orb}$
as a function of primary mass ($m_{\rm 1}$, left panel) and mass ratio
($q=m_{2}/m_{1}$, right panel).

\begin{figure*}
\centering
\includegraphics[width=17cm,clip,angle=0]{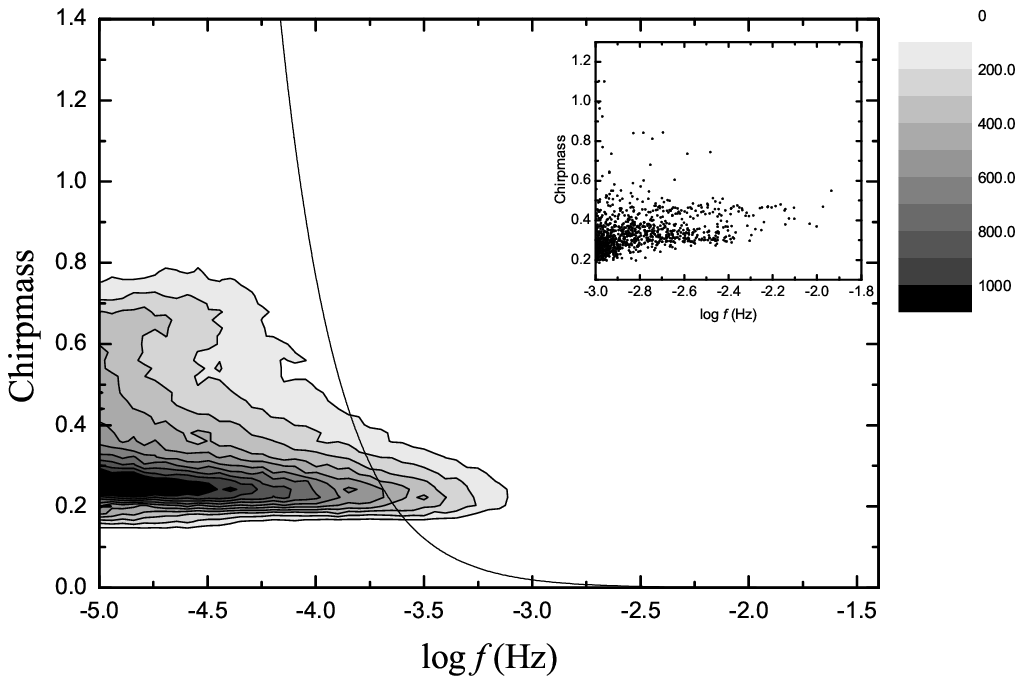}
\caption{Distribution of the chirp mass and frequency of DWDs. 
The contour scale to the right of the figure shows the number of
systems in each bin.
The present DWD population is plotted, but the total number of DWDs 
($2.76\times10^{8}$) is reduced by a factor of 100. 
The solid line denotes the boundary for DWDs which will merge within 
15 Gyr. The upper-right panel shows the distribution
of chirp mass at high GW frequency $\log f~> -3.0$ Hz. Bin sizes are
$\Delta \log f = 0.05 $, and $\Delta \mathcal{M} = 0.02 {\rm M_{\odot}}$. } 
\label{fig_chirpmass}
\end{figure*}

\subsection{Chirp masses}
\label{sec_chirp}

We plot the number density distribution of DWD chirp masses
  against frequency in Fig.~\ref{fig_chirpmass}. Using
Eq.~\ref{eq_grtime} and assuming 15 Gyr for
the age of universe, we obtain a relation between chirp mass and a
critical frequency 
for DWDs with a circular orbit, {\it i.e.} $e=0.0$,
\begin{equation}
\mathcal{M}=3.05\times10^{-7}f_{\rm c}^{-8/5}.
\end{equation}
Figure~\ref{fig_chirpmass} shows this relation; 
DWDs with  $f > f_{\rm c}$ will merge within 15 Gyr.

We can understand the distribution of chirp mass from the view 
of stellar evolution. A star only develops a degenerate 
helium core to become a helium WD either if its main sequence 
progenitor evolves to the giant branch and loses its hydrogen envelope 
{\it prior} to core helium ignition, or if the hydrogen-envelope
is removed during main-sequence evolution so that the star fails to
reach the giant branch and evolves directly from either the main sequence or the
Hertzsprung gap to the He WD cooling track. 
The mass of the He WD will be $\lesssim 0.5 {\rm M_{\odot}}$, 
depending somewhat on metallicity. 

If the mass of the He core is $\gtrsim 0.5 {\rm M_{\odot}}$,
core helium burning will be ignited. Following core-helium exhaustion, 
the star will evolve to become a degenerate carbon-oxygen (CO) 
or oxygen-neon-magnesium (ONeMg) WD with mass in the range of $\approx 0.5 -  
1.44 {\rm M_{\odot}}$ following an asymptotic giant branch, 
naked helium giant or hot subdwarf phase, 
depending on the chemical composition
and mass of the progenitor. 

With the population synthesis parameters adopted here, Figure
  \ref{fig_chirpmass} shows a principal ridge in chirp mass at 
$\mathcal{M}\approx 0.3 {\rm M_{\odot}}$. Figure \ref{fig_type}
  demonstrates that this ridge corresponds to the He+He DWDs and
  represents the largest population of DWDs in our model. 
A smaller and broader ridge is located at 
$\mathcal{M}\approx 0.65 {\rm M_{\odot}}$ and corresponds to the CO+CO
DWDs (Fig. \ref{fig_type}). CO+He DWDs have a range of chirp masses 
intermediate between the He+He and CO+CO DWS (Fig. \ref{fig_type}),
making the chirp-mass distribution in Fig.~\ref{fig_chirpmass} appear 
relatively continuous. 

\begin{figure*}
\centering
\includegraphics[width=16cm,clip,angle=0]{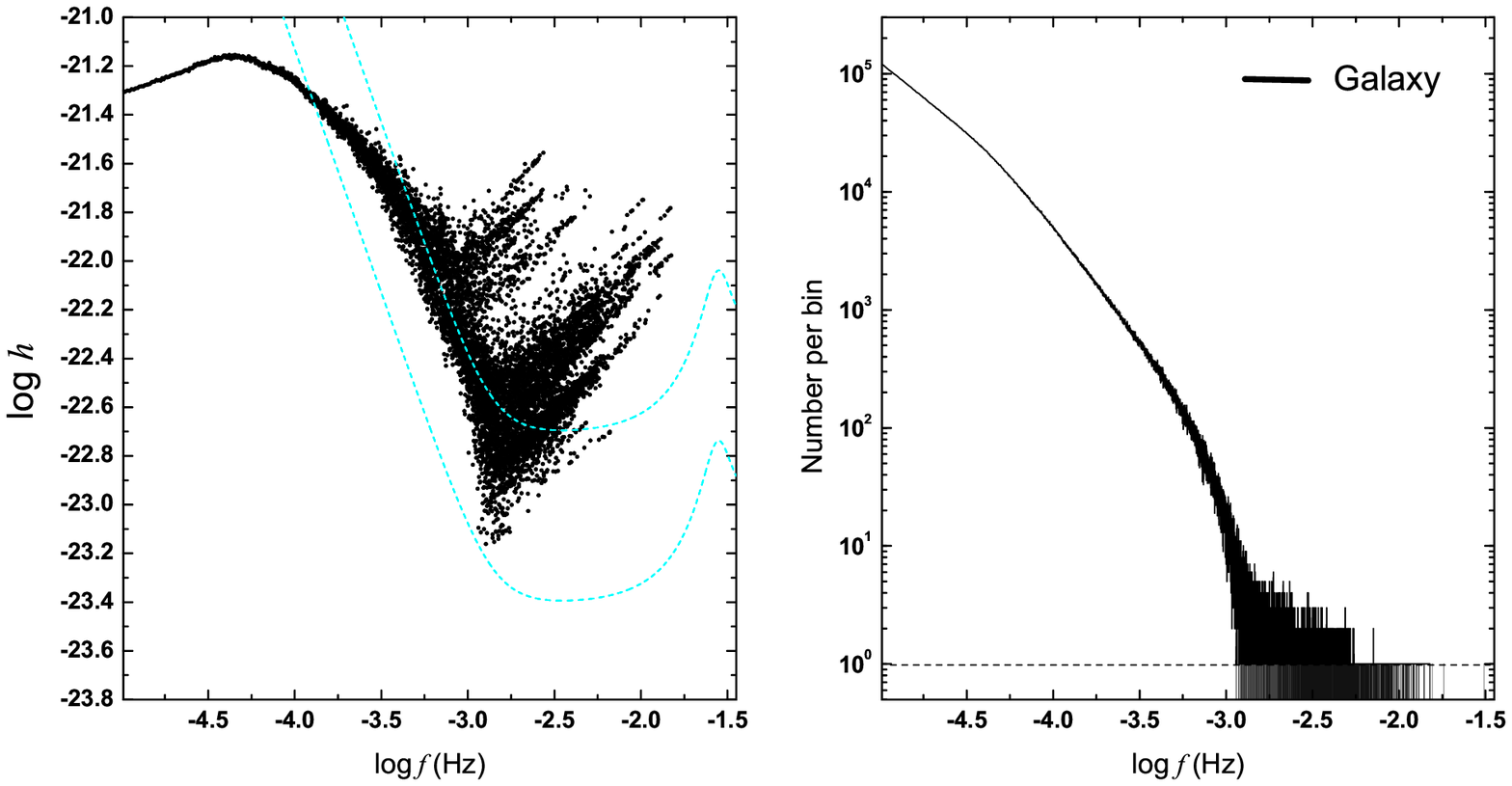}
\caption{The number of DWDs per frequency bin (right
  panel) and their GW signal (left panel). In the left panel, $\log h$
represents the strain amplitude. Colored dashed lines indicate the
expected LISA sensitivity for a S/N = 1 (lower) or 5 (upper). The bin
size $\Delta f = 3.17\times10^{-8} {\rm Hz}$.}
\label{fig_totals}
\end{figure*}

\subsection{The total GW amplitude spectrum from DWDs and 
comparison with LISA}
\label{sec_totalgw}

In order to synthesize the amplitude spectrum for DWDs, 
we choose a LISA integration time $T =
3.16\times10^{7}$\,s (1 yr), so that the frequency resolution $\Delta
f = 1/T = 3.17\times10^{-8}\,{\rm Hz}$, which we adopt for the
size of the frequency bin. Hereafter, GW frequencies $f$ are given in
Hz, unless otherwise stated.

Some frequency bins in the high-frequency domain will contain an
individual source if the bin is small enough, arising from a rapid
decrease in the number of Galactic binaries towards shorter periods.
These are the so-called {\it resolved sources}. \citet{Evans87}
discussed the relation between the detector bandwidth $\Delta f$ and
integration time $T$. They plotted the amplitude of DWDs with
integrations times of $10^{6}$\,s and $10^{8}$\,s, and found that
all binaries with a frequency above $\sim3$\,mHz should be
resolved if the integration time is $>10^{8}$\,s.

Figure~\ref{fig_totals} illustrates the predicted total number of DWDs
per frequency bin (right panel) and the GW signal (left)
these DWDs would produce. Some bins in the range $-3.0 < \log f < -1.80$ only
contain a single system (Resolved DWDs), making up $\sim$0.012\%
(33670, Table \ref{tab_birthrates}) of the total number of DWDs.

LISA is designed to be a space-based GW detector, consisting of 
3 satellites flying in formation to form a Michelson interferometer 
with an arm length of 5$\times10^{6}$ km. Noise arises mainly from the laser 
tracking system (position noise) and parasitic forces on the proof mass 
of an accelerometer (acceleration noise) \citep{Larson00}. We can convert 
the noise signal to an equivalent GW signal in frequency 
space by  
\begin{equation}
h_{\rm f}=2\sqrt{\frac{S_{\rm n}}{R}},
\end{equation}
where $S_{\rm n}$ is the total strain noise spectral density, 
$h_{\rm f}$ is the root spectral density and $R$ is the GW 
transfer function given by \citet{Larson00}. 

For a continuous monochromatic source, such as a DWD with a circular 
orbit, which is observed over a time $T$, the root spectral density 
will appear in a Fourier spectrum as a single spectral line 
in the form \citep{Larson00}
\begin{equation}
h_{\rm f}=\frac{h}{\sqrt{\Delta f}}=h\sqrt{T}.
\end{equation}
So, for an observation time $T = 1$ yr, 
the root spectral density $h_{\rm f}=5.62\times10^{3}h$.

\begin{figure*}
\centering
\includegraphics[width=18cm,clip,angle=0]{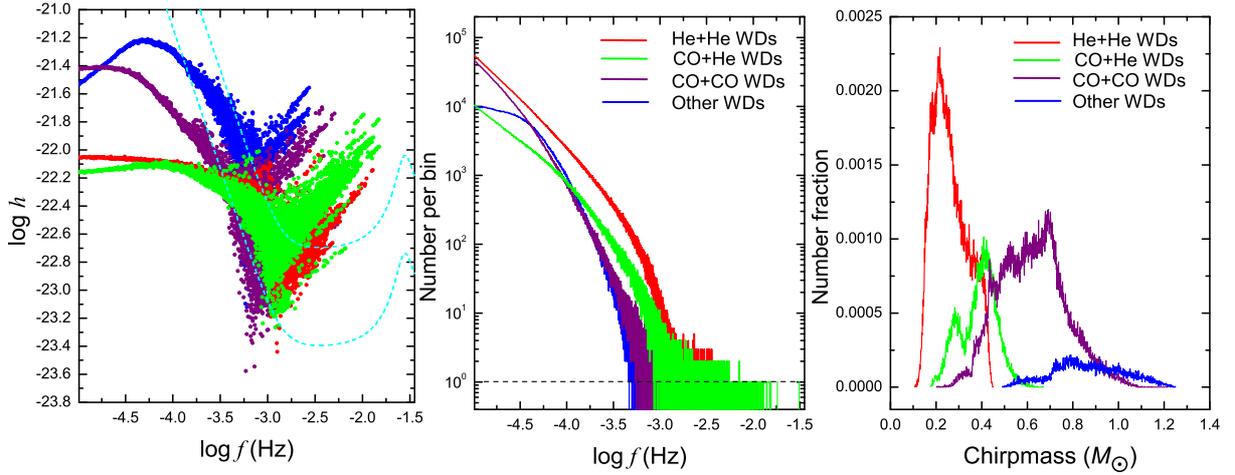}
\caption{The GW signal due to different types of DWD. $\log h$
  represents the strain amplitude (left panel).  The number
  distribution is shown in the middle panel. The right panel
  illustrates the distribution of chirp mass due to different DWD
  types, normalized by the total number (2.76$\times10^{8}$) of 
  DWDs in the Galaxy (bin size $\Delta \mathcal{M}=0.001 {\rm M_{\odot}}$).
  The LISA sensitivity and frequency bins are as in
Fig.~\ref{fig_totals}. } \label{fig_type}
\end{figure*}

\begin{figure*}
\centering
\includegraphics[width=16cm,clip,angle=0]{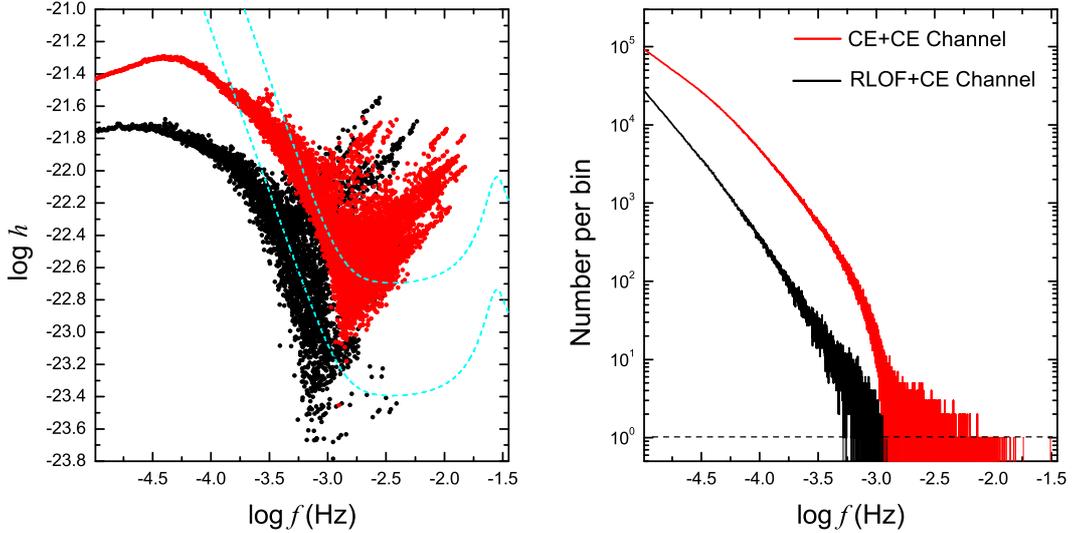}
\caption{The GW signal due to DWDs from different formation channels.
The left panel presents the strain amplitudes. The
number distribution is shown in the right.
The LISA sensitivity and frequency bins are as in
Fig.~\ref{fig_totals}. Note that the exposed core + CE channel is a very 
minor channel ($<3\%$) and has almost no influence on the GW signal 
at frequencies $\log f> -5$ ($P_{\rm orb}< 55.56$\,hr).} \label{fig_channel}
\end{figure*}

\begin{figure*}
\centering
\includegraphics[width=16cm,clip,angle=0]{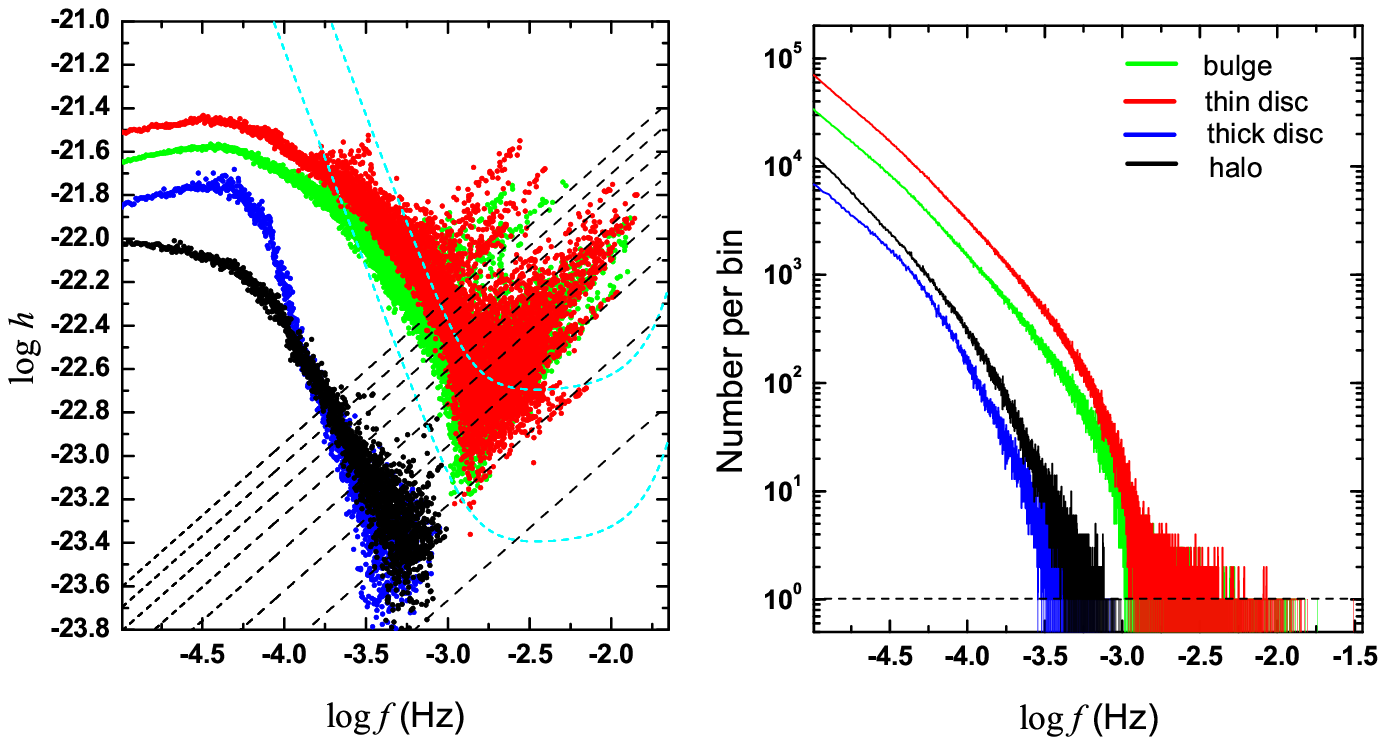}
\caption{The GW signal due to DWDs from different populations (see key).
The black dashed lines denote the relation of amplitude and frequency
from Eq.~\ref{eq_logh} at chirp masses 0.12, 0.22, 0.32, 0.42, 0.52,
0.62, 0.72 and 0.82 $M_{\odot}$ (from bottom). 
The LISA sensitivity and frequency bins are as in
Fig.~\ref{fig_totals}.}
\label{fig_fractions}
\end{figure*}

To demonstrate the detectability of the predicted GW
signals due to DWDs, Figs.~\ref{fig_totals} -- \ref{fig_thindisc} 
show the expected LISA sensitivity for S/N=1 and 5.

\subsection{The GW signal from various DWD types and channels}
\label{sec_typechan}

Figure \ref{fig_type} shows the contribution to the GW signal 
from different types of DWD in our model, 
including those containing two helium WDs (He+He),
a carbon-oxygen WD and a helium WD (CO+He), and two CO WDs
(CO+CO). DWDs containing at least one ONeMg white
dwarf are designated ``ONeMg''.

Figure \ref{fig_type} shows that CO+CO and ONeMg DWDs 
have a stronger total strain amplitude for $\log f < -3.5$ ($P_{\rm orb} >
0.9$\,hr), up to almost 0.6 dex higher than that of He+He and CO+He DWDs,
despite being less numerous. Since the strain amplitude for a single DWD of
given frequency is proportional to $\mathcal{M}^{5/3}$
(Eq.~\ref{eq_hne}), the larger chirp masses of the CO+CO and ONeMg
DWDs dominate the numerical superiority of the He+He DWDs.

CO+CO and ONeMg DWDs are formed in more massive progenitor
binaries than He+He and CO+He DWD progenitors. Such binaries
undergo RLOF or CE ejection when the stars are physically larger,
and hence are in longer period systems and less likely to form
short-period DWDs. Consequently the slope of the number-frequency
distribution for CO+CO and ONeMg DWDs is steeper than for
less-massive DWDs, and the nett contribution to the strain amplitude
falls away more quickly. 

Consequently, for $\log f > -3.5$, the signals from all four DWD types converge and
then at $\log f > -3.0$, the strain amplitude appears to scale roughly as
the average chirp mass for the DWD type. The main reason is that the number of
systems per frequency bin is here very small and generally $\leq 3$.
The contribution from ONeMg DWDs remains highest because of the
chirp-mass effect. However, our model predicts negligibly few ONeMg DWDs at 
$\log f > -2.5$, and negligibly few CO+CO DWDs at $\log f > -2.25$ ($P_{\rm
  orb}<6.65$ min), so these contributions vanish at these frequencies.

Fig. \ref{fig_type} shows that CO+He DWDs are the dominant 
GW source at very high frequency ($\log f > -2.3$). 
This may be understood as a consequence of evolutionary 
age. In general, ONeMg DWDs and CO+CO DWDs correspond to larger progenitor 
masses and hence a smaller number of progenitors, 
faster evolution and a lower detection probability. 
On the other hand, He WDs have larger radii, so He+He DWDs tend to 
merge more quickly after a double CE phase.

Figure \ref{fig_channel} compares the contribution of different DWD
formation channels. The stable RLOF+CE and CE+CE channels dominate the
production of DWDs, with the CE+CE channel generating 66.7\% of all
DWDs and dominating the GW signal at all frequencies $\log f > -5$
except in the range $-3.0 < \log f < -2.3$ where the
contribution from the RLOF+CE and CE+CE channels may be comparable.
The exposed core + CE channel is a very minor channel ($<3\%$) 
and has almost no influence on the GW signal at frequencies $\log f> -5$
($P_{\rm orb}< 55.56$\,hr) in our current model, but this would depend 
on a tidally enhanced stellar wind as discussed by \citet{Han98} (see
also \S \ref{sec_formation}). 

Figure \ref{fig_channel} also shows that the highest-frequency DWDs
  come exclusively from the double CE channel. This channel is expected to be 
the most important channel for close DWDs because a deep spiral-in
phase is necessary to release sufficient orbital energy to eject a
common envelope, finally leading to the formation of  a very close binary. 

In Table \ref{tab_type}, we list the total numbers and relative
contributions of various types of DWD and evolutionary channels, as
well as their current birth rate and the potential number of resolved
systems.  We find that CO+He and He+He DWDs would be the most significant
GW sources at
$\log f> -2.92$ ($P_{\rm orb}< 27.7$\, min),
at which frequency resolved sources emerge from our sample.  We
however cannot neglect the population ``ONeMg'' as they also have a
considerable resolved number. CE+CE is the dominant formation channel for
resolved DWDs, producing 3.8 times more systems than the stable
RLOF+CE channel.

\begin{table*}
\caption{Birth rates of DWDs with various types and different
  formation channels. $\nu$ = current birth rate, $N$ = total
  number, \% = percentage of the type of DWDs making up the total resolved DWDs
  in the Galaxy, N$_{\rm Resolved}$ = number of resolved DWDs. We
  take the thin disc model with $Z=0.02$.}
\label{tab_type}
\begin{center}
\begin{tabular}{lccccccc}
\hline
 &$\nu$ & $N$ & $N/N_{\rm gal}$ & $N_{\rm Resolved}$ &  \% \\
 \hline
 types of WD binaries\\
 He+He & $1.34\times10^{-2}$ &$1.06\times10^{8}$& 38.41\% & 19936 &59.21\%\\
 CO+He & $5.07\times10^{-3}$ &$4.11\times10^{7}$& 14.89\% & 12852 &38.17\%\\
 CO+CO & $1.15\times10^{-2}$ &$1.08\times10^{8}$& 39.13\% &  586 &1.74\%\\
 ONeMg & $2.13\times10^{-3}$ &$2.09\times10^{7}$& 7.57\%  & 296 &0.88\%\\
 \hline
 formation channels\\
 RLOF+CE&$9.86\times10^{-3}$ &$8.27\times10^{7}$& 29.96\% &  1061 &3.15\%\\
 CE+CE & $2.12\times10^{-2}$ &$1.84\times10^{8}$& 66.67\% & 32609 &96.85\%\\
 other channels & $1.04\times10^{-3}$ &$9.3\times10^{6}$& 3.37\% & 0 &0\\
 \hline
 Total & $3.21\times10^{-2}$ &$2.76\times10^{8}$&         & 33670 & \\
 \hline
 \hline
\end{tabular}
\end{center}
\end{table*}

\subsection{The GW signal from various Galactic populations}
\label{sec_diffpop}

Figure~\ref{fig_fractions} shows the GW amplitude (right panel) and
the number of DWDs per frequency bin (left panel) due
to each component of the Galaxy, including bulge, thin disc, thick
disc and halo. This figure demonstrates that DWDs in the thin disc
and bulge generate a strong strain amplitude which should be
observable by LISA. DWDs in the halo and thick disc might
make a substantially smaller contribution to the GW amplitude and only
at $\log f < -3$. These DWDs will not contribute to the LISA
background.

Figure~\ref{fig_fractions} shows that the main difference 
in the GW signal between different populations is at very high
frequencies. At lower GW frequencies the 
discrepancy is principally caused by the number of DWDs. 
At very high frequency, the GW signal strongly depends on the star formation 
history. We note that a persistent quasi-exponential SFR was applied to the 
bulge and thin disc,  but only a single star burst was applied at the beginning of the 
thick disc and halo. Figure~\ref{fig_fractions} also shows that 
there is only a slight difference between the signal from the bulge and 
thin disc, despite having a differnt IMF and spatial distribution.

Figures \ref{appfig} and \ref{fig_fractions} show that distance 
has a only small effect on the DWD signal 
at high frequency. Therefore, if we assume $d\approx10$ 
kpc, take logarithms of Equation~\ref{eq_hneco}, and combine
with the relation between orbital period and GW frequency ($f=2/P_{\rm
  orb}$: circular orbits), we obtain 
\begin{equation}
\log~h=-20.13+\frac{5}{3}\log~\left(\frac{\mathcal{M}}{\rm
M_{\odot}}\right)+\frac{2}{3}\log~f.
\label{eq_logh}
\end{equation}
 We plot this relation in Fig.~\ref{fig_fractions} as black dashed
lines, which demonstrates the effect of chirp mass on
GW amplitude. Thus at high frequency, where individual DWDs can be
resolved, chirp mass will strongly influence the strength of the GW
signal. It will be used to compare our results with others in \S\,\ref{s_disc}.

\begin{figure*}
\centering
\includegraphics[width=14cm,clip,angle=0]{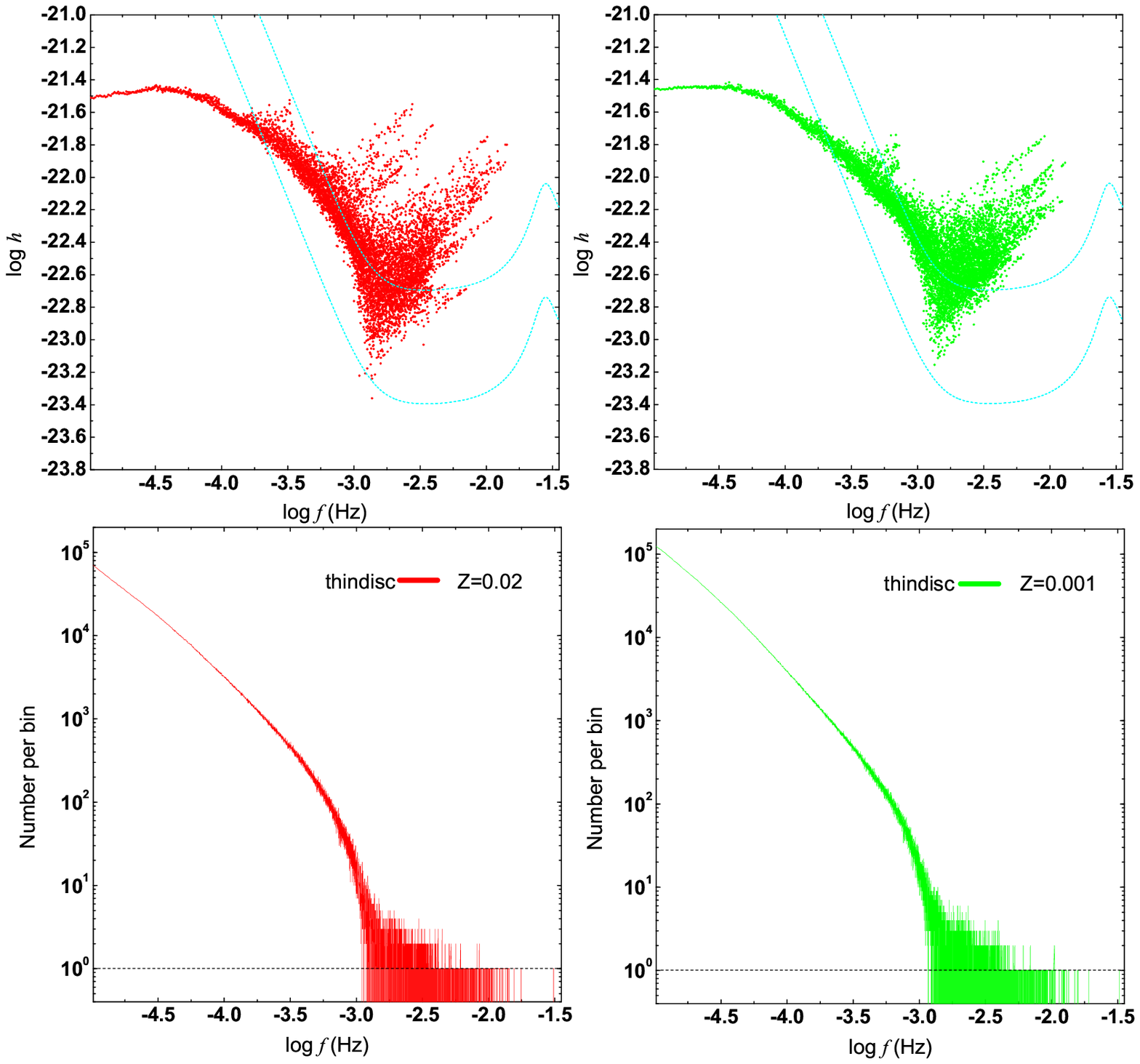}
\caption{The GW signal obtained for different thin disc models. The
red line is for $Z=$0.02, green for $Z=$0.001. The LISA sensitivity
and frequency bins are as in Fig.~\ref{fig_totals}.}
\label{fig_thindisc}
\end{figure*}

\subsection{The effect of metallicity}
\label{sec_metal}

There is a large number of parameters in our model for
  synthesizing the evolution of large numbers of stars and, since
  every synthesis takes a large amount of computing time, too many to
  explore individually in this paper. However, by far the most
  important parameter in any stellar evolution model, after mass, is
  metallicity. This will therefore be a crucial factor in determining 
  the properties of the DWD population \citep{Pols98,Han98,Hurley00}. 

In order to show the influence of metallicity 
on the GW signal, we performed an additional computation for the thin disc 
using identical model parameters except for the metallicity. We have compared 
metallicities $Z=0.02$ and $0.001$.

Figure~\ref{fig_thindisc} demonstrates that the shape of the
GW spectrum is slightly affected by metallicity at low frequency,
but is not much affected at high frequency. A lower
metallicity increases both the total number of DWDs, and the number of
resolved DWDs. In our thin disc model, these numbers are
$1.7\times10^8 (2.3\times10^8)$ and 18870 (22670) for
$Z=0.02 (0.001)$, respectively.

The difference in the shape of the  spectrum arises because
at lower metallicities, asymptotic-giant branch stars are able to
develop more massive cores, and also achieve higher luminosities and
larger radii. This leads to a greater number of more massive
WDs in longer initial period orbits, and also to more
DWDs surviving.

\subsection{Detached, semi-detached and merged DWDs}

Following formation of a DWD, evolution under the influence of GW may
follow one of several paths: \\
1) If the initial binary is too wide, it will not merge within a
Hubble time or equivalent; we adopt
10\,Gyr. Considering only the $1.7\times10^8$ thin disc DWDs (Table~\ref{tab_birthrates})
this fraction represents some 94\% (or $1.6\times10^8$) of the total.\\
2) Of the remaining 6\% that will evolve into contact within 10 Gyr,
the majority, 95\% or  $9.6\times10^6$ DWDs are formed
with $q>q_{\rm c}$ and will merge. \\
3) The remainder, having $q<q_{\rm c}$ will undergo stable Roche
lobe overflow. This will decrease
their mass ratios further, their orbits will widen,
and the systems will only stay in contact by further GW radiation.
These correspond to some of the AM\,CVn systems\footnote{
\citet{Nelemans01a} also discuss AM\,CVn systems which are formed from
low-mass helium stars with degenerate companions.}.
Their number is $\approx 5\%$ of those that evolve into contact, or
$4.8\times10^5$ DWDs, or $0.3\%$ of all DWDs formed in the thin
disc\footnote{
These figures correspond to model I of \citet{Nelemans01a}, in which
there is no tidal coupling between the accretor spin and the orbital angular
momentum. \citet{Nelemans01a} find that with effective tidal
coupling, the number of AM\,CVn systems may increase by two orders of
magnitude (their model II). It is not the intention of this paper to discuss AM\,CVn
systems in detail.}.

\subsection{Resolved sources of GW radiation}

\citet{Evans87} suggested that degenerate dwarf binaries would be
promising sources of GW and pointed out that the detection of GW is
related to the integration time. By computing a large number of DWDs,
\citet{Nelemans01b} concluded that wide DWDs dominate the GW signal
at $\log f \lesssim -3.4$, and that AM\,CVn systems could be resolved
at $\log f \gtrsim -2.8$, but would produce a lower signal due to their
small number. Subsequently, \citet{Nelemans04} showed that AM\,CVn systems
would be good candidates for LISA and that $\approx 11000$ AM\,CVn
systems could be resolved. 
By ''resolved'', we mean that, over a certain observation time $T$, 
a frequency bin ($\Delta f=1/T$) contains only one binary.

In general, our methods and results are similar.
One difference is that, by including their contribution, 
we confirm that DWDs from the halo and thick disk
would make little contribution to the LISA signal.
A second difference is that, by adopting a different IMF,
star formation rate and total mass for the bulge and the thin disc,
we obtain an increase in the total number of
compact DWDs at high frequencies, so that the number of resolved
sources increases slightly.

Our results confirm that all DWDs with frequency $\log f > -2.25$ would be resolved
with a LISA integration time equal to 1 yr.
In \S\,3.3, we noted that resolved systems with $\log f >
-3.0$ would account for 0.012\% of the total number of DWDs,
of which 29\% will become semi-detached.
These numbers agree reasonably  with previous results,
although \citet{Nelemans01b,Nelemans04} found
$\approx 50\%$ to be semi-detached.

\S \ref{sec_typechan} points out that most resolved DWDs should
have formed from the CE+CE channel and should be dominated by CO+He
and He+He DWDs. They should be the most significant resolved sources for
LISA due to their number and strain amplitude and would
dominate the LISA GW signal at $\log f> -2.25$, ($P_{\rm orb} < 6.65$\,min).
It may also be possible to infer their type (CO+He or He+He) from
their chirp masses (Eq.~\ref{eq_chirpmass}) by determining $P_{\rm
  orb}$ and $\dot{P}_{\rm orb}$, possibly from
optical observations.

CO$+$CO and ONeMg DWDs have a stronger strain amplitude at
$\log f< -3.12$ ($P_{\rm orb}> 0.732$\,hr), resulting from their
larger chirp masses. For frequencies $-2.25 > \log f > -3.12$,
the signal from all four types of DWD would overlap.

\begin{table*}
\caption{Main parameters and their value in our simulation. See Table 
\ref{tab_densitylaws} for the Galactic structure parameters.
See \S \ref{sec_model} for an explanation of the parameters. }
\label{tab_mainparameter}
\begin{center}
\begin{tabular}{lcccccc}
\hline
stellar evolution parameters
&$\gamma$ & $\alpha_{\rm RLOF}$ & $B$ & $\beta_{\rm w}$ & $v_{\rm w}$  \\
&1.5      &             0.5    & 1000 & 1.5            &  20  \\
\hline
\hline
 & bulge & thin disc & thick disc & halo \\
\hline
age (Gyr)& 10  & 10  & 11 & 14  \\
current $M_{\rm g}$ ($M_{\odot}$) & 2$\times10^{10}$   & 5.2$\times10^{10}$  & 2.6$\times10^{9}$ & 1$\times10^{9}$ \\
current SFR ($M_{\odot}\rm yr^{-1}$)  & 1.45 & 3.37  & single burst  & single burst\\
IMF     &  $\propto m^{-2.35}$   &  Eq. \ref{eq_imftn}  & $\propto m^{-1.5}$ & $\propto m^{-1.5}$  \\
$Z $    &   0.02   &  0.02  & 0.02 & 0.001  \\
$q $     &  flat   &  flat  & flat & flat  \\
$a $     &  Eq. \ref{eq_a}   &   Eq. \ref{eq_a}  & Eq. \ref{eq_a} & Eq. \ref{eq_a}  \\
$P_{\rm e} $    &  $\propto 2e$   &  $\propto 2e$   & $\propto 2e$  & $\propto 2e$   \\
\hline \hline
\end{tabular}
\end{center}
\end{table*}

\begin{table}
\caption{Estimated errors in DWD birthrates for the thin disk}
\label{tab_error}
\begin{center}
\begin{tabular}{lccrr}
\hline
$p$ & $\langle p\rangle$ & $\langle\delta p\rangle$
                         & $\partial \nu / \partial p$ & $\delta \nu_p / \nu$ \\
\hline
$\gamma            $ & $ 1.5$ & 0.1 & $3.4\times10^{-4}$ & 0.016 \\
$\log Z            $ & $-1.7$ & 0.5 & $4.3\times10^{-3}$ & 0.102 \\
$t\,{\rm Gyr^{-1}} $ & $14$   & 1   & $0.1\times10^{-3}$ & 0.005 \\
$\alpha_{\rm IMF}$  & $-2.2$ &  0.1  & $8.6\times10^{-4}$ & 0.041 \\
$a_{\rm SFR}$  &  11   &   1  & $1.0\times10^{-3}$ & 0.047  \\
\hline \hline
\end{tabular}
\end{center}
\end{table}

\subsection{Error Analysis}
\label{error}

Modelling stellar populations and hence estimating the numbers,
properties and distributions of stars in the Galaxy is subject to 
several sources of error (or uncertainty). These may na\"ively be divided
into statistical errors arising from the implementation of the Monte
Carlo process, and systematic errors arising from choices made for
the model parameters (of which over thirty may be identified in \S\,2;
principal values are indicated in Table \ref{tab_mainparameter}).

The Monte Carlo procedures involve two stages, the first of which
creates a sample of $\int n dt' = 1.84 \times 10^4$ DWDs in the thin disc for example,
on which subsequent samples are interpolated. Assuming Poisson statistics,
the error ($1 \sigma$) in this first number is $\approx \sqrt{\int n dt'} \approx 1.4\times10^2$
or $\approx 0.8\%$. This represents a lower limit on the error in subsequent
quantities, including total birth rates, merger rates and numbers of
DWDs. The statistical error on the number distributions and strain
amplitude spectra will be at least as large as this, increasing as
$(N(f)df)^{-0.5}$  towards higher frequencies.

In order to investigate some of the systematic errors, we 
have carried out reduced simulations for the thin disc
population with an initial sample of $10^{5}$ binaries and by varying
each of the five most important parameters (Table \ref{tab_error}). 
Whilst we omit other parameters from this exploration ($B$, for example)
to save computation time, their systematics have been discussed by others \citep{Han98}.

Systematic errors can be estimated by considering the gradient
$\partial \nu / \partial p$ of the birthrate $\nu$ with respect
to a model parameter $p$, and multiplying by an expectation value
$\langle\delta p\rangle$ for the uncertainty in $p$.
Considering $\gamma$, metallicity (here expressed as $\log Z$),
age $t$, IMF, and SFR to be representative of inputs with large
uncertainty, fractional errors
$\delta \nu_p / \nu \equiv \partial \nu / \partial p \times \langle \delta
p \rangle / \nu $
are indicated in Table~\ref{tab_error}. To simplify,
we have here assumed the disk IMF takes the simpler form
$\xi(m) \approx m^{\alpha}$ and the disk SFR takes the form  $a (\exp(t-t_0) /
\tau)+b(t-t_0)$, so that they may be characterised by $\alpha_{\rm IMF}$ and
$a_{\rm SFR}$ respectively; $\tau$ is constrained by the current SFR.
$\langle \delta \alpha_{\rm IMF} \rangle \approx 0.1$
is suggested by comparing \citet{Kroupa93} with \citet{Kroupa01},
while $\langle \delta a_{\rm SFR} \rangle \approx 1$ is taken from
the uncertainty in the thin disc mass given by \citet{Klypin02}.

The dominant uncertainties in $\nu$ arise from the metallicity
approximation $Z=0.02$, at least in the thin disc, and the SFR.
Again making a na\"{\i}ve approximation that all $p$ are
independent, adding the error contributions quadratically gives a
total relative error ($1\sigma$) from these sources
$\delta \nu / \nu \approx 0.34$.

These numbers are also useful for comparing results between
different models by substituting the uncertainty $\delta p$ with the
difference between adopted values $\Delta p$.

In our model the Galactic DWD birthrate
$\nu-\xi \approx 3.2\times10^{-2} ~{\rm yr^{-1}}$
(Table \ref{tab_birthrates}).
\citet{Han98} found a similar value
($3\times 10^{-2} ~{\rm yr^{-1}}$).
For the thin disc alone we find the DWD birthrate to be
$2.1\times10^{-2} ~{\rm  yr^{-1}}$
while \citet{Nelemans01b} obtained
$2.5\times10^{-2} ~{\rm yr^{-1}}$.
The slight difference between our model and
that of \citet{Nelemans01b} can be understood primarily
in terms of the difference $\Delta a_{\rm SFR}=3$ and $\Delta \alpha_{\rm IMF}=0.3$

The birth rate of supernovae Ia (0.0013 yr$^{-1}$) in
our model is slightly smaller than that of \citet{Han98}
(0.003 yr$^{-1}$) and \citet{Nelemans01b} (0.002 yr$^{-1}$),
and is probably also a consequence of the IMF and SFR. The fact 
that we set a limit to the Eddington accretion rate (Eq. \ref{eddington}) 
restricts the SN Ia rate \citep{Livio00}.

Another effect on the birth rate and number arises from
tidally-enhanced stellar winds and / or wind-driven mass transfer,
which is related to a coefficient $B$ (\S \ref{windandaccretion}).
\citet{Han98} finds that the birth rate and number vary slightly with $B$.
We adopt $B=1000$ following \citet{Han98}, but much larger values
might be justified; with $B=10000$ \citep{Tout91}, the mass-loss rate
could be 150 times larger than the Reimers' rate when the star
nearly fills its Roche lobe.

\section{Discussion}
\label{s_disc}

In addition to work by
\citet{Nelemans01b,Nelemans04} already discussed, the question of GW radiation from DWDs
and their contribution to the LISA signal has also been addressed by
 \citet{Webbink98}, \citet{Hils00}, \citet{Hiscock00}, \citet{Willems07}, \citet{Liu09}
and \citet{Ruiter09}, amongst others.

In our model, the GW strain amplitude from the total Galactic
DWD population would be $\approx 2.51\times10^{-22} -
6.31\times10^{-24}\,{\rm Hz^{1/2}}$  for $\log f > -4.0$,
and $\approx 5.01\times10^{-22} - 5.01\times10^{-23}\,{\rm Hz^{1/2}}$
for $\log f < -4.0$. The larger strain at lower
frequencies is primarily due to the larger number of DWDs per unit
frequency.

At $\log f > -3$, the GW signal from DWDs in the bulge and thin disc combined
equals the total Galactic GW signal.  The thick disc and halo
make no contribution (Fig.~\ref{fig_fractions}) because all DWDs
that formed in these components that will merge have already done so.
At lower frequencies ($\log f \leq -3$),
\citet{Ruiter09} found that the halo signal at $\log f=-3$, for example,
$h_{\rm f}=10^{-20}\,{\rm Hz^{1/2}} ( \equiv \log h\,{\rm Hz^{-1/2}} =
  -23.75$ )
is a factor $\approx10$ lower than that of the disc+bulge, where
$h_{\rm f}=10^{-19}\,{\rm Hz^{1/2}} ( \equiv \log h\,{\rm Hz^{-1/2}} =
  -22.75$).
They estimated there to be $1.5\times10^{9}$ halo white dwarfs
with a total mass of $10^{9}\,{\rm M_{\odot}}$ in a halo with
$Z=0.0001$ and age 13 Gyr.
At $\log f = -3.3$, we have $<\log h>\approx-23.3$, roughly a factor three
stronger than \citet{Ruiter09}, but in a halo with $Z=0.001$ and age
14 Gyr.

Another difference between our model and the model of \citet{Ruiter09}
arises from the chirp masses ({\it cf.} \S~\ref{sec_diffpop}).
Figure \ref{fig_fractions} shows the strain-amplitude relation as a
function of chirp masses; since there are only one or two systems in 
the majority of high frequency bins, it is possible to infer that the chirp 
masses of most DWDs in our model are greater than $0.1\,{\rm
  M_{\odot}}$. In comparison, the chirp masses of high-frequency DWDs 
in the model of \citet{Ruiter09} appear to be less than $0.1\,{\rm
  M_{\odot}}$.
The different distribution of chirpmass in the two models may be due to 
differences in $Z$ and age adopted in each case. 
Our results more closely resemble those of \citet{Nelemans01b}.

The GW radiation from AM\,CVn binaries and helium cataclysmic
variables was calculated by \citet{Hils00}, who found that these two
populations only provide a slight enhancement on the confusion noise
for LISA below 3\,mHz and even no increase at higher frequency.
\citet{Nelemans01b} and \citet{Webbink98} showed a confused background
GW signal due to DWDs which is similar to ours at low frequency, but
different at high frequency.

\citet{Hiscock00} found that the background signal from halo WD
binaries could be five times stronger than the expected contribution from
Galactic disk binaries by assuming that the fraction of WDs in
binaries is the same in the halo as in the disk. This suggested the
possibility of a GW signal from halo DWDs although probably the
assumption overestimated the number of DWDs in halo, which would
justify further investigation.  \citet{Willems07} considered GW
radiation from eccentric DWDs formed through interactions in globular
clusters and found LISA could provide unique dynamical identifications
of these systems. This requires further investigation.

Since the change of GW frequency (or orbital period) $\dot{f}$ can
be measured by electromagnetic observations or LISA, $\mathcal{M}$
is given by
\begin{equation}
\dot{f}=5.8\times10^{-7}(\mathcal{M}/{\rm M}_{\odot})^{5/3}f ^{11/3}
~~ {\rm s^{-2}}. \label{eq_chirpmass}
\end{equation}
Hence, for a small number of high-frequency systems likely to be
detected by LISA, it will be possible to determine $\mathcal{M}$
directly.
By knowing $\mathcal{M}$ and $P$, we can calculate the distance
$d$ from Eq.~\ref{eq_hne}, which is vital for constraining the
distribution of DWDs and obtaining a better understanding of the
structure of the Galaxy.

In this paper, we have not considered the BH in the center of the
bulge or low-mass BHs elsewhere in the Galaxy. More attention should be
paid to these in future work, as they could greatly interfere with
the ability of LISA to detect resolved systems \citep{Nelemans01b}.

\section{Conclusion}

A self-consistent numerical simulation of the gravitational wave
signal from the entire population of double white dwarf binaries
(DWDs) in the Galaxy has been carried out. We have used a population
synthesis approach to follow the stellar evolution, and a
comprehensive Galactic model in which we suppose that the Galaxy is
composed of a bulge, thin disc, thick disc, and halo. This model
demonstrates a significant contribution to the GW signal due to
DWDs.

We have discussed the birth rates, local densities and total numbers 
of DWDs in
each fraction of the Galaxy, as well as of supernovae Ia, noting
that these are affected by the particular choice of IMF, SFR,
metallicity and age adopted in our model (Table~\ref{tab_birthrates}).

The GW signals from different components of the Galaxy have been
computed, as well as the contribution from various types of DWD and
from DWDs formed by different formation channels. The corresponding
birth rates and numbers have been given (Table~\ref{tab_type}).

Formation channels CE$+$CE and stable RLOF$+$CE play the most
important role to produce DWD populations detectable
by LISA at $\log f > -4.5$ ($P_{\rm orb} < 17.57$ h).
DWDs with the shortest orbital periods in our model come from the
CE$+$CE channel. The Exposed Core$+$CE channel is an almost negligible
channel for the formation of DWDs.

We find that CO$+$He and He$+$He white dwarf pairs  dominate the GW signal
at high frequency ($\log f> -2.3$), while CO$+$CO and ONeMg DWDs
make the major contribution at  $\log f < -2.3$.

We find that $\approx 33\,000$ DWDs could be resolved by LISA.  Most would have formed through
the CE$+$CE channel, and the model favours the majority being CO$+$He
and He$+$He DWDs. Ground-based observations might determine which by
measuring their chirp masses.

\begin{acknowledgements}
The Armagh Observatory is supported by a grant from the Northern
Ireland Dept. of Culture Arts and Leisure. SY thanks the Leverhulme 
Trust for support. SY and CSJ thank the referee for his/her constructive 
comments and also thank G. Nelemans and Z. Han for the discussion. 
\end{acknowledgements}

\bibliographystyle{aa}
\bibliography{g14827yu}

\end{document}